\documentclass[aps,prb,amsmath,twocolumn,superscriptaddress,showpacs,preprintnumbers,reprint]{revtex4-1}
\pdfoutput=1
\usepackage{graphicx}
\newcommand{\cunipa}{Cu$_5$-NIPA}
\newcommand{\eff}{\text{eff}}
\newcommand{\AFM}{\text{AFM}}
\newcommand{\FM}{\text{FM}}
\newcommand{\onehalf}{\left|\frac12\right.\rangle}

\newcommand{\threehalf}{\left|\frac32\right.\rangle}
\newcommand{\fivehalf}{\left|\frac52\right.\rangle}

\usepackage{color}

\usepackage[colorlinks,bookmarks=false,citecolor=blue,linkcolor=red,urlcolor=blue]{hyperref}
\usepackage{verbatim}

\def\be{\begin{equation}}
\def\ee{\end{equation}}
\def\bea{\begin{eqnarray}}
\def\eea{\end{eqnarray}}

\def\vec{\mathbf}

\def\mc{\mathcal}

\begin{document}

\title{Magnetization and spin dynamics of the spin $S=\frac{1}{2}$ hourglass nanomagnet Cu$_{5}$(OH)$_{2}$(NIPA)$_{4}\cdot$10H$_2$O}

\author{R. Nath}
\affiliation{School of Physics, Indian Institute of Science 
Education and Research, Thiruvananthapuram-695016, India}

\author{A. A. Tsirlin}
\email{altsirlin@gmail.com}
\affiliation{Max Planck Institut f\"{u}r Chemische Physik fester
Stoffe, N\"{o}thnitzer Str.~40, 01187 Dresden, Germany}
\affiliation{National Institute of Chemical Physics and Biophysics, 12618 Tallinn, Estonia}

\author{P. Khuntia}
\affiliation{Max Planck Institut f\"{u}r Chemische Physik fester
Stoffe, N\"{o}thnitzer Str.~40, 01187 Dresden, Germany}

\author{O.~Janson}
\affiliation{Max Planck Institut f\"{u}r Chemische Physik fester
Stoffe, N\"{o}thnitzer Str.~40, 01187 Dresden, Germany}
\affiliation{National Institute of Chemical Physics and Biophysics, 12618 Tallinn, Estonia}

\author{T.~F\"orster}
\affiliation{Max Planck Institut f\"{u}r Chemische Physik fester
Stoffe, N\"{o}thnitzer Str.~40, 01187 Dresden, Germany}
\affiliation{Dresden High Magnetic Field Laboratory, Helmholtz-Zentrum Dresden-Rossendorf, 01314 Dresden, Germany}

\author{M.~Padmanabhan}
\affiliation{School of Chemistry, Indian Institute of Science 
Education and Research, Thiruvananthapuram-695016, India}

\author{J. Li}
\affiliation{Department of Chemistry $\&$ Chemical Biology, Rutgers University, Piscataway, NJ 08854, USA}

\author{Yu.~Skourski}
\affiliation{Dresden High Magnetic Field Laboratory, Helmholtz-Zentrum Dresden-Rossendorf, 01314 Dresden, Germany}

\author{M. Baenitz}
\affiliation{Max Planck Institut f\"{u}r Chemische Physik fester
Stoffe, N\"{o}thnitzer Str.~40, 01187 Dresden, Germany}

\author{H. Rosner}
\affiliation{Max Planck Institut f\"{u}r Chemische Physik fester
Stoffe, N\"{o}thnitzer Str.~40, 01187 Dresden, Germany}

\author{I. Rousochatzakis}
\email{i.rousochatzakis@ifw-dresden.de}
\affiliation{Institute for Theoretical Solid State Physics, IFW Dresden, 01171 Dresden, Germany}


\begin{abstract}
We report a combined experimental and theoretical study of the spin $S\!=\!\frac{1}{2}$ nanomagnet Cu$_{5}$(OH)$_{2}$(NIPA)$_{4}\cdot$10H$_2$O (Cu$_5$-NIPA). Using thermodynamic, electron spin resonance and $^1$H nuclear magnetic resonance measurements on one hand, and {\it ab initio} density-functional band-structure calculations, exact diagonalizations and a strong coupling theory on the other, we derive a microscopic magnetic model of Cu$_5$-NIPA and characterize the spin dynamics of this system. The elementary five-fold Cu$^{2+}$ unit features an hourglass structure of two corner-sharing scalene triangles related by inversion symmetry. Our microscopic Heisenberg model comprises one ferromagnetic and two antiferromagnetic exchange couplings in each triangle, stabilizing a single spin $S\!=\!\frac12$ doublet ground state (GS), with an exactly vanishing zero-field splitting (by Kramer's theorem), and a very large excitation gap of $\Delta\!\simeq\!68$ K. Thus, Cu$_5$-NIPA is a good candidate for achieving long electronic spin relaxation ($T_1$) and coherence ($T_2$) times at low temperatures, in analogy to other nanomagnets with low-spin GS's. Of particular interest is the strongly inhomogeneous distribution of the GS magnetic moment over the five Cu$^{2+}$ spins. This is a purely quantum-mechanical effect since, despite the non-frustrated nature of the magnetic couplings, the GS is far from the classical collinear ferrimagnetic configuration. Finally, \cunipa\ is a rare example of a $S\!=\!\frac12$ nanomagnet showing an enhancement in the nuclear spin-lattice relaxation rate $1/T_1$ at intermediate temperatures.
\end{abstract}


\maketitle

\section{\textbf{Introduction}}
The field of molecular nanomagnets has enjoyed an enormous experimental and theoretical activity over the last few decades.\cite{kahn1990,*gatteschi2006} Owing to the nanoscopic size of their elementary magnetic units, these compounds provide experimental access to a plethora of quantum mechanical (QM) effects, including quantum tunneling of the magnetization\cite{gunther1995,thomas1996} or the N\'eel vector,\cite{Loss98} quantum phase interference,\cite{wernsdorfer1999}  
level-crossings and magnetization plateaux.\cite{Taft1994,*Julien1999} They also allow to probe on the macroscopic scale  
the crossover from quantum to classical physics.\cite{Stamp96,*Swarzschild97} Finally, molecular magnets are promising materials for spintronic applications\cite{bogani2008} and quantum computing.\cite{leuenberger2001}

Here we report on the magnetic behavior, microscopic magnetic model, and spin dynamics of Cu$_5$(OH)$_2$(NIPA)$_4\cdot 10$H$_2$O, hereinafter referred to as Cu$_5$-NIPA, where the acronym NIPA stands for the \mbox{5-nitro-isophtalic} acid ligand. It is 
an hourglass-shaped molecular magnet comprising five Cu$^{2+}$ spin-$\frac12$ ions. The ground state (GS) of this magnet has a low spin value $S\!=\!\frac12$ with a very large spin gap of $\Delta\!\simeq\!68$ K. Therefore, \cunipa\ behaves as a rigid spin $S\!=\!\frac12$ entity in a wide temperature range, and resembles other \mbox{spin-$\frac12$} molecular magnets, such as V$_{6}$.\cite{ioannis2005,luban2002} Given that both compounds comprise $s\!=\!\frac12$ spins, we expect similarly long electron spin-phonon relaxation times $T_1$, which allow for the observation of rich hysteresis effects in pulsed fields.\cite{chiorescu2000,ioannis2005} However, in contrast to V$_{6}$, here we do not expect abrupt steps in the magnetization curve,\cite{ioannis2005} because the present compound features an odd number of half-integer spins, thus the zero-field splitting vanishes exactly by Kramer's theorem. 
    
In analogy with other molecular magnets with low-spin GS's, such as iron-sulfur clusters,\cite{Dilg99,Shergill91,Guigliarelli95}
heterometallic rings,\cite{Ardavan07} iron trimers,\cite{Mitrikas08} V$_{15}$,\cite{bertaina08} and single-molecule magnets (SMM),\cite{Schlegel08} we expect that \cunipa\ manifests also long coherence times $T_2$, which is a crucial step towards implementations in quantum computing.\cite{Stolze04}

Another attractive feature of Cu$_5$-NIPA is the presence of two corner-sharing scalene triangles, related by inversion symmetry.\cite{zhao2011,liu2011} The spin triangle is the most elementary unit for highly frustrated magnetism,\cite{balents2010} 
while its chirality may induce a finite magnetoelectric coupling,\cite{trif2008,*Bulaevskii2008,kamiya2012} and may also be used as a qubit.\cite{GeorgeotMila2010} In molecular magnetism, the spin triangle is common in many compounds,\cite{kahn1990,*gatteschi2006} such as V$_{15}$,\cite{chiorescu2000,chaboussant2002,*furukawa2007}  
Cu$_3$ clusters,\cite{choi2006,*choi2008} the chiral Dy$_3$ cluster,\cite{Dy3chirality}
the cuboctahedron Cu$_{12}$La$_8$,\cite{cu12a,*cu12b} as well as the giant icosidodecahedral keplerates Mo$_{72}$Fe$_{30}$,\cite{Fe30} W$_{72}$Fe$_{30}$,\cite{WFe30} Mo$_{72}$Cr$_{30}$,\cite{Cr30} Mo$_{72}$V$_{30}$\cite{botar2005,*V30b} 
and W$_{72}$V$_{30}$,\cite{WV30} which host highly frustrating kagome-like physics.\cite{ioannis08}

\begin{figure*}
\includegraphics{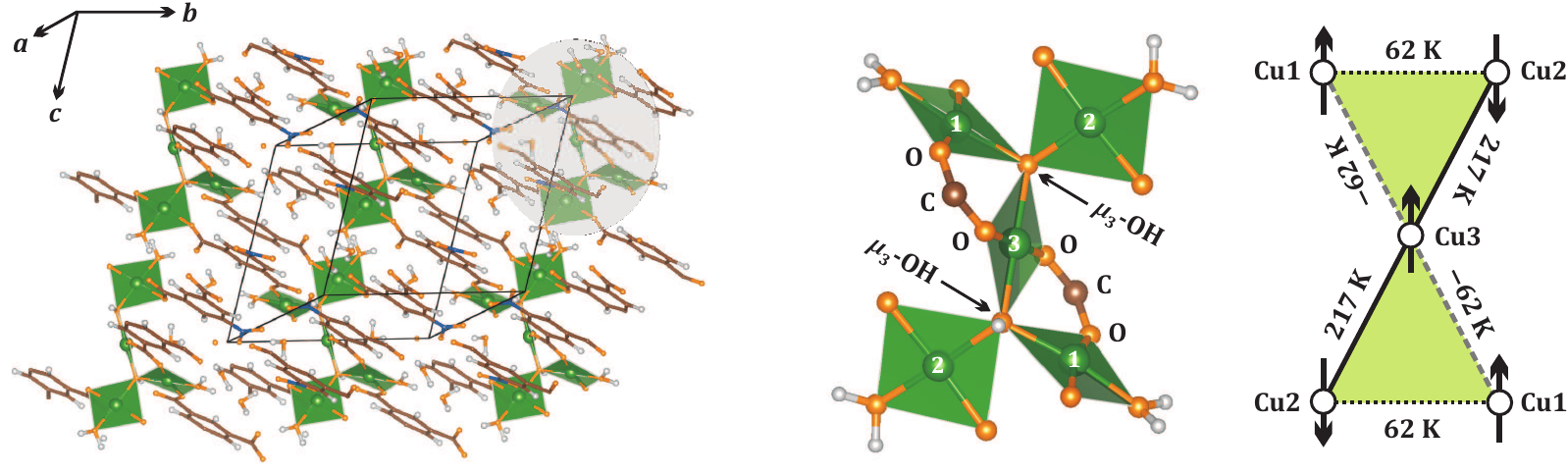}
\caption{\label{fig:structure}
(Color online) Left panel: crystal structure of \cunipa\ showing the stacking of the Cu$_5$ units (one unit is highlighted by gray shading) through the NIPA ligand molecules, into a rigid three-dimensional metal-organic framework, with the shortest distance of 8.6~\r A between the magnetic units. Middle and right panels: structure of the Cu$_5$ magnetic unit and relevant magnetic interactions according to the fit of the experimental susceptibility data (Fig.~\ref{fig:chi-fit}). The numbers $1-3$ denote the crystallographic positions Cu1--Cu3, with Cu3 residing at the inversion center. Arrows show the classical ferrimagnetic ground state.}
\end{figure*}

The distinct feature of the present compound is the strong distortion of its regular spin triangles,  
featuring three nonequivalent Cu positions, denoted by Cu1, Cu2, and Cu3 (see Fig.~\ref{fig:structure}). 
Each spin triangle is centered by the $\mu_3$-OH group\footnote{Here, $\mu_3$ means that the central oxygen atom is connected to three Cu atoms.} 
providing Cu--O--Cu superexchange pathways, while the carboxyl (COO$^-$) group of the NIPA ligand creates an additional Cu1--Cu3 pathway (see Fig.~\ref{fig:structure}, middle). This topology leads to three drastically different interactions, one of which is ferromagnetic (FM). Despite the fact that these couplings do not compete with each other classically, 
the GS has strong QM character, which is partly reflected in a strongly inhomogeneous distribution of the magnetic moment over the five Cu$^{2+}$ spins. 

Of particular interest is our experimental finding of an enhancement of the $^1$H nuclear spin-lattice relaxation rate $1/T_1$ at 
a characteristic temperature slightly below the spin gap ($T\simeq 40$ K). While such an enhancement has been repeatedly found in numerous antiferromagnetic (AFM) homometallic\cite{baek2004} and heterometallic\cite{amiri2010} rings of spins $s>\frac12$, it is very rare for $s\!=\!\frac{1}{2}$. We argue that the origin of the peak is the same in both cases, namely the slowing down of the phonon-driven magnetization fluctuations. \cite{baek2004,santini2005,ioannisPRB2007,ioannisPRB2009} However, there are several qualitative differences with the case of homometallic rings, related to the sparse excitation spectrum of \cunipa\ and the presence of inequivalent Cu$^{2+}$ sites.

The organization of this article is the following. We begin in Sec.~\ref{sec:Methods} with details on the sample preparation and methods. Next, we discuss bulk specific heat and magnetization (Sec.~\ref{sec:thermo}), as well as local electron spin resonance (Sec.~\ref{sec:esr}) and $^1$H nuclear magnetic resonance data (Sec.~\ref{sec:nmr}).   
The microscopic description of \cunipa\ in terms of the isotropic Heisenberg model is based on {\it ab initio} density-functional band-structure calculations (Sec.~\ref{sec:model}) and facilitates the evaluation of magnetic properties (local magnetizations, nature of the GS, etc.) using exact diagonalizations (Sec.~\ref{sec:ED}) and a strong coupling theory (Sec.~\ref{sec:SCE}). 
Finally, we conclude with a discussion and some interesting perspectives of this study in Sec.~\ref{sec:discussion}.

\section{\textbf{Methods}}\label{sec:Methods}
A powder sample of \cunipa\ was prepared according to the procedure described in Ref.~\onlinecite{zhao2011}. Sample purity was checked by powder x-ray diffraction (Huber G670 Guinier Camera, CuK$_{\alpha1}$ radiation, $2\theta=3-100^{\circ}$ angular range).

The magnetic susceptibility ($\chi$) was measured in the temperature range \mbox{$1.8$ K $\leq $ $T$ $\leq $ $400$ K} in an applied field of $\mu_0H=1$~T using the commercial Quantum Design MPMS SQUID. The magnetization isotherm $M$ vs. $H$ was measured at $T=2$~K in fields up to 14~T using the vibrating sample magnetometer (VSM) option of Quantum Design PPMS. Additionally, pulsed-field measurements in fields up to 60~T were performed at 1.4~K in the Dresden High Magnetic Field Laboratory. Details of the measurement procedure are described in Ref.~\onlinecite{tsirlin2009}. 

The heat capacity $C_{\rm p}(T)$ was measured on a small piece of pellet over the temperature range \mbox{$2$ K $\leq $ $T$ $\leq $ $300$ K} in zero field using the Quantum Design PPMS.

The electron spin resonance (ESR) measurements were performed at Q-band frequencies ($f=34$~GHz) using a standard spectrometer together with a He-flow cryostat that allows us to vary the temperature from 1.6 to 300~K. ESR probes the absorbed power $P$ of a transversal magnetic microwave field as a function of a static and external magnetic field $B$. To improve the signal-to-noise ratio, we used a lock-in technique by modulating the static field, which yields the derivative of the resonance signal $dP/dB$.

\begin{figure*}
\includegraphics{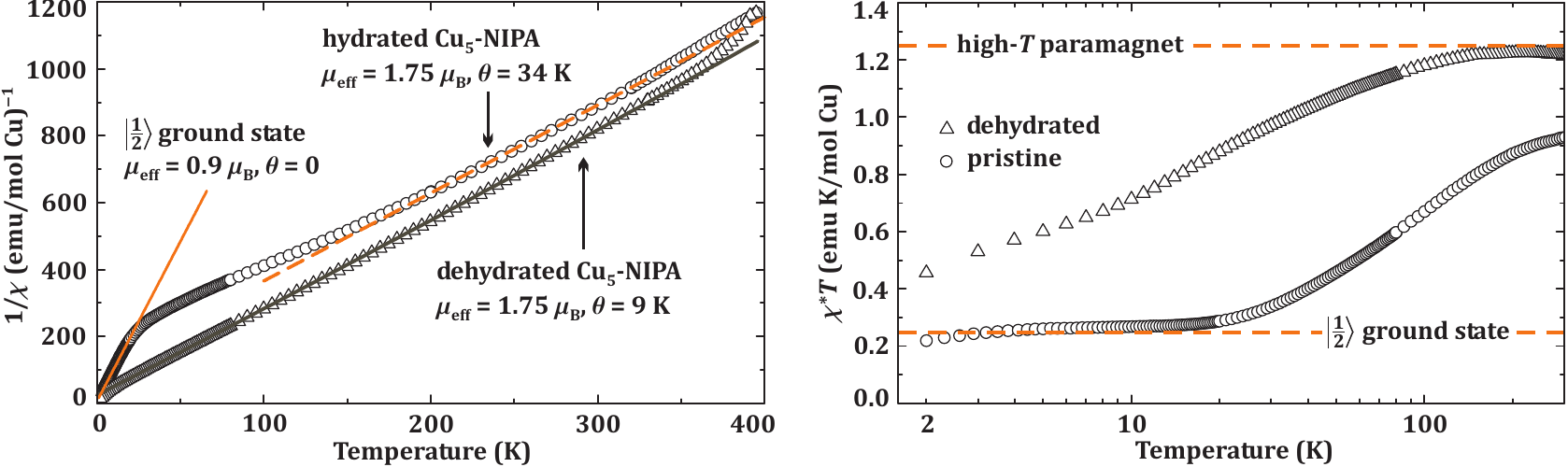}
\caption{\label{fig:chi}
(Color online) Left panel: Inverse magnetic susceptibility of Cu$_5$-NIPA measured upon heating from 2~K (circles, hydrated form) 
and upon cooling from 400~K (triangles, dehydrated form). Lines show Curie-Weiss fits to Eq.~\eqref{eq:cw}, as described in the text. Right panel: $\chi^*T$ plot showing the formation of the $\onehalf$ state in the pristine sample ($\chi^*T=\frac14$) and the high-temperature paramagnetic regime of the dehydrated sample ($\chi^*T=\frac54$). Here, $\chi^*=\chi\left(\frac{Ng^2\mu_B^2}{k_B}\right)^{-1}$, $N=N_A/5$, and $g=2.2$ according to ESR (Sec.~\ref{sec:esr}). For the $\chi(T)$ plot, see Fig.~\ref{fig:chi-fit}.}
\end{figure*}

The nuclear magnetic resonance (NMR) measurements were carried out using pulsed NMR technique on $^{1}$H nuclei (nuclear spin $I=\frac12$ and gyromagnetic ratio $\gamma_{n}/2\pi = 42.576$~MHz/T) in the $2-230$~K temperature range. The NMR measurements were performed at two different radio frequencies of 70~MHz and 38.5~MHz, which correspond to an applied field of about $1.6608$~T and $0.9135$~T, respectively. Spectra were obtained by Fourier transform of the NMR echo signal. The $^{1}$H spin-lattice relaxation rate, $1/T_{1}$, was measured by using the conventional saturation pulse sequence.

The electronic structure of \cunipa\ was calculated in the framework of density functional theory (DFT) using the \texttt{VASP} code\cite{vasp1,*vasp2} for crystal structure optimization and \texttt{FPLO}\cite{fplo} for the evaluation of magnetic parameters. The generalized gradient approximation (GGA)\cite{pbe96} exchange-correlation potential was augmented with the mean-field DFT+$U$ correction for Coulomb correlations in the Cu $3d$ shell. The DFT+$U$ parameters were chosen as $U_d=9.5$~eV (on-site Coulomb repulsion), $J_d=1$~eV (on-site Hund's exchange), and fully-localized-limit (FLL) flavor of the double-counting correction, following earlier studies of Cu$^{2+}$-based magnets.\cite{janson2012,*tsirlin2010} Reference calculations with other exchange-correlation potentials and double-counting corrections arrived at qualitatively similar results. The reciprocal space was sampled by a $k$ mesh with 64 points in the first Brillouin zone. The convergence with respect to the $k$ mesh was carefully checked. Details of the computational procedure are reported in Sec.~\ref{sec:model}.

Thermodynamic and GS properties of the Cu$_5$ molecule were evaluated by full numerical diagonalizations.

\section{Experimental results}

\subsection{Thermodynamic properties}\label{sec:thermo}
\subsubsection{Magnetization}
\label{sec:chi}
The magnetic susceptibility of \cunipa\ was measured for the as-prepared sample under field-cooling (FC) condition upon heating from 1.8~K to 400~K. At 400~K, the sample was kept inside the SQUID magnetometer for about 20~minutes, and $\chi(T)$ was measured again upon cooling from 400~K to 1.8~K. The drastic difference between the heating and cooling curves (Fig.~\ref{fig:chi}) indicates the decomposition of \cunipa\ shortly above room temperature. Indeed, the sample color changed from blue to green, and the weight loss of 13\,\% was detected. This weight loss is in good agreement with the preceding thermogravimetric data of Ref.~\onlinecite{liu2011} that reported the weight loss of 13.6\,\% at $380-400$~K. Although the decomposition process is tentatively ascribed to the release of water molecules (expected weight loss of 13.15\,\%)\cite{liu2011}, no detailed information on the decomposed sample is available in the literature.

The pristine sample shows a sharp decrease in the magnetic susceptibility (increase in $1/\chi$) upon heating and a bend around 30~K followed by the nearly linear regime of $1/\chi$ above $200-250$~K. In contrast, the decomposed sample remains paramagnetic down to at least 10~K. Below 320~K, where the decomposition process is finished, the inverse susceptibility of dehydrated \cunipa\ follows a straight line that can be fitted with the Curie-Weiss law in the $20-320$~K temperature range,\footnote{Note that we do not use the temperature-independent term $\chi_0$, because fits for the pristine sample are done in a narrow temperature range that does not support the evaluation of three variable parameters. The lack of $\chi_0$ results in subtle deviations of the high-temperature effective moment from the $g$-value obtained in the model fit (Fig.~\ref{fig:chi-fit}).}
\begin{equation}
\chi=\dfrac{C}{T+\theta}.
\label{eq:cw}
\end{equation}
The resulting Curie constant $C\simeq 0.383$~emu~K/(mol~Cu) leads to an effective moment $\mu_{\eff}\simeq 1.75$~$\mu_B$, which is close to 1.73~$\mu_B$ expected for spin-$\frac12$. The Weiss temperature is $\theta\simeq 9$~K. 

The inverse susceptibility of the pristine \cunipa\ sample does not have a well-defined paramagnetic region, because the data above $320-350$~K are affected by the decomposition. A tentative Curie-Weiss fit in the $220-320$~K range yields an effective moment of $\mu_{\eff}\simeq 1.75$~$\mu_B$ ($C=0.383$~emu~K/(mol~Cu)), which is same as in the dehydrated sample. However, the Weiss temperature of $\theta\simeq 34$~K in the pristine sample is notably larger than $\theta\simeq 9$~K observed in the dehydrated sample. 

The conspicuous difference between the hydrated and dehydrated samples suggests a huge effect of dehydration on the magnetism of the Cu$_5$ molecule. The dehydration proceeds in a single step and results in the loss of all 10 water molecules per formula unit. As 4 out of these 10 molecules enter the first coordination sphere of Cu (Fig.~\ref{fig:structure}, left), a large effect on the local environment of Cu sites and, thus, on the magnetism should be expected.\cite{[{For a similar example, see: }][{}]schmitt2009} Unfortunately, no structural data for the dehydrated version of \cunipa\ are presently available. Therefore, we restrict ourselves to a detailed study of the hydrated compound. 

\begin{figure}
\includegraphics{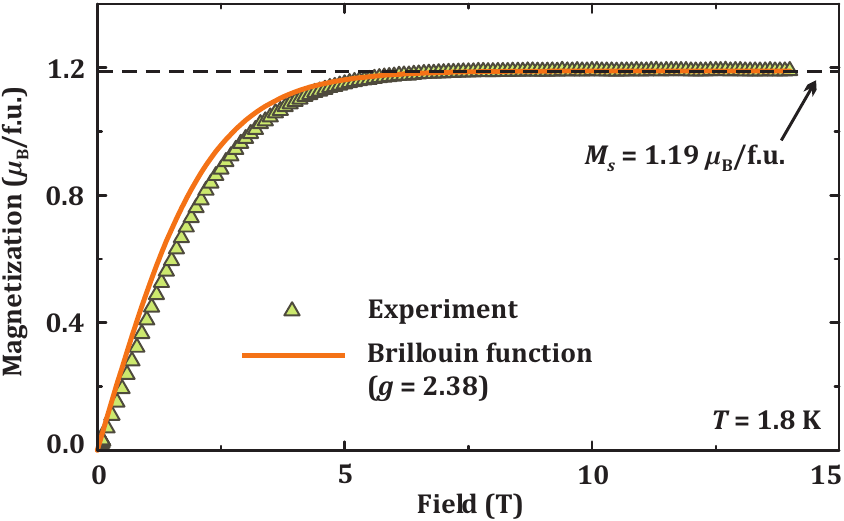}
\caption{\label{fig:mvsh} 
(Color online) Magnetization curve of Cu$_5$-NIPA measured at 1.8~K up to 14~T. The dashed line denotes the saturation of the $\onehalf$ state at $M_s=1.19$~$\mu_B$/f.u. The solid line is the Brillouin function calculated at 1.8~K with $g=2.38$.}
\end{figure}

At low temperatures, the inverse susceptibility of \cunipa\ approaches another linear regime, with the Curie constant $C=0.101$~emu~K/(mol~Cu) ($\mu_{\eff}\simeq 0.90$~$\mu_B$) and vanishingly small $\theta\simeq 0.5$~K. This low-temperature paramagnetic state can be also seen in the $\chi^*T$ plot ($\chi^*=\chi\left(Ng^2\mu_B^2/k_B\right)^{-1}$ is the reduced susceptibility), where $\chi^*T$ approaches the value of $\frac14$ expected for the total spin $S=\frac12$ per molecule (Fig.~\ref{fig:chi}, right). Likewise, the magnetization curve of \cunipa\ (Fig.~\ref{fig:mvsh}) saturates at $M_s=1.19$~$\mu_B$/molecule, which is, however, higher than the free-electron value of 1~$\mu_B$/f.u. This difference can be well accounted for by the large $g$-value of 2.38 according to $M_s=gS\mu_B$. The same $g$-value explains the low-temperature effective moment: $C=Ng^2\mu_B^2/3k_B=0.106$~emu~K/(mol~Cu), where we use $N=N_A/5$ for the magnetic moment of $\frac12$ per Cu$_5$ molecule. As we explain later, this change in the $g$-value (Sec.~\ref{sec:ED}) can be traced back to the presence of three non-equivalent Cu positions with different $g$-tensor anisotropies. At room temperature, the Cu spins are nearly independent, and their $g$-values average with same weights for all Cu positions. At low temperatures, the powder averaging of the $g$-values is determined by the distribution of the magnetization in the $S=\frac12$ GS.

Our results suggest that at low temperatures \cunipa\ is in the paramagnetic state with the total moment of $S=\frac12$ per Cu$_5$ molecule. This state is further denoted as $\onehalf$. The experimental magnetization curve follows the Brillouin function 
\begin{equation}
  B(H)=g\mu_BS\times\text{th}\left(\frac{g\mu_BSH}{k_BT}\right)
\end{equation}
with $S=\frac12$, $g=2.38$, and $T=1.8$~K (Fig.~\ref{fig:mvsh}). A marginal departure of the experimental curve toward higher fields may be due to anisotropies and/or intermolecular couplings. The maximum deviation of about 0.3~T puts an upper limit of about 0.5~K on the total energy of such couplings. This low energy scale is in agreement with the large spatial separation between the molecules (Fig.~\ref{fig:structure}, right).

Below 4~K, the susceptibility of \cunipa\ departs from $\chi^*T=\frac14$ (Fig.~\ref{fig:chi}, right). This decrease in $\chi^*T$ may indicate an evolution of the system toward a long-range magnetic order between the Cu$_5$ molecules. However, we were unable to see any clear signatures of a magnetic transition in the susceptibility data measured down to 2~K.

\subsubsection{Specific heat}
\begin{figure}
\includegraphics[]{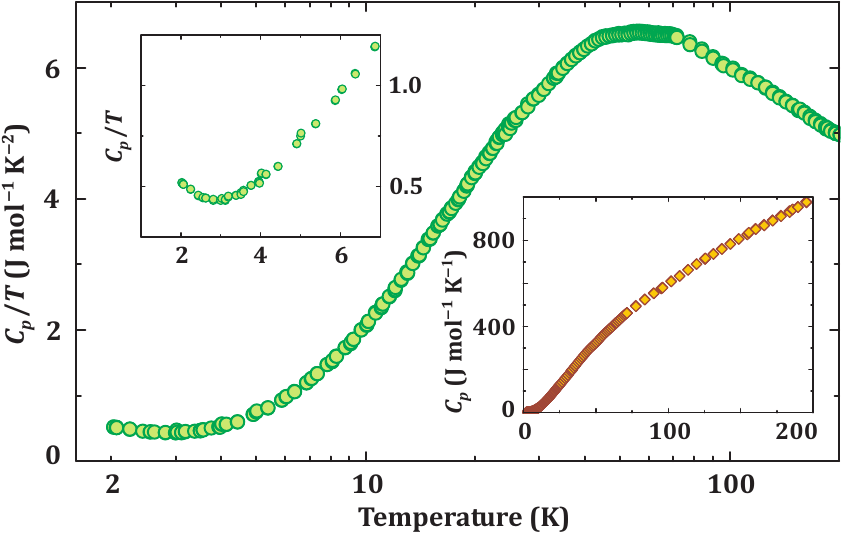}
\caption{\label{fig:heat} (Color online) Specific heat of \cunipa\ divided by temperature ($C_p/T$). The inset in the upper left corner magnifies the data at low temperatures. The inset in the bottom right corner shows the specific heat ($C_p$).}
\end{figure}
The specific heat ($C_p$) of \cunipa\ is smooth down to 2~K (Fig.~\ref{fig:heat}). It steeply increases up to $60-70$~K and shows a slower increase at higher temperatures, as shown by a broad maximum in the temperature dependence of $C_p/T$. The signal is dominated by the phonon contribution, whereas the magnetic part is quite small. The total magnetic entropy of \cunipa\ is $5R\ln 2\simeq 28.8$~J~mol$^{-1}$~K$^{-1}$, which is only 2~\% of the total entropy of about 1100~J~mol$^{-1}$~K$^{-1}$ released up to 200~K (the latter is obtained by integrating the temperature dependence of $C_p/T$). 

At 200~K, the heat capacity of \cunipa\ is still very far from its maximal Dulong-Petit value of $C_p=3RN\simeq 2767$~J~mol$^{-1}$~K$^{-1}$, because \cunipa\ features a large number of high-energy phonon modes related to the \mbox{O--H}, C--C, C--N, and C--O vibrations. The entropy associated with these modes can be released at very high temperatures, only. The complexity of the \cunipa\ structure and the presence of multiple phonon modes of different nature hinder a quantitative analysis of the specific heat data. A non-magnetic reference compound would be ideal to extract the magnetic contribution and analyze it in more detail. Unfortunately, such a reference compound is presently not available.

At low temperatures, the heat capacity of \cunipa\ does not fall smoothly to zero. The temperature dependence of $C_p/T$ shows an increase below 3~K (see the upper left inset of Fig.~\ref{fig:chi}). The origin of this behavior is presently unclear. The increase in $C_p/T$ could signify a Schottky anomaly or a proximity to the magnetic ordering transition. However, our susceptibility data, as well as NMR (Sec.~\ref{sec:nmr}), rule out any magnetic transition down to 2~K. 
The heat capacity data are consistent with these observations.

\subsection{ESR}
\label{sec:esr}
\begin{figure}[!t]
\includegraphics{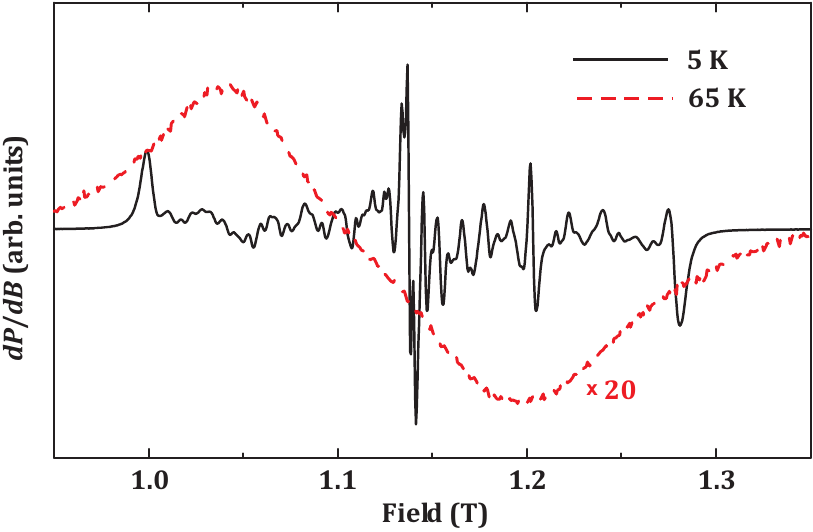}
\caption{\label{fig:esr1}
(Color online) Typical ESR signal of \cunipa\ at 5~K (solid line) and 65~K (dashed line). For better visibility, the intensity of the high-temperature signal was increased by a factor of 20.
}
\end{figure}

The ESR spectrum of \cunipa\ could be detected from the lowest measured temperature (5~K) up to 135~K. Above 135~K, the ESR line broadens and eventually becomes invisible. Between 70~K and 135~K, the spectra contain only one line that can be fitted with a single Lorentzian function (1L-Fit) providing the ESR parameters, the linewidth $\Delta B$ and $g$-factor, $g=\frac{h\nu}{\mu_B B_{\text{res}}}$, where $B_{\text{res}}$ is the resonance field. Below 70~K, the single ESR line splits, so that at $40-60$~K the fit is only possible with two Lorentzian lines (2L-Fit), whereas at even lower temperatures a plethora of narrow lines appear (Fig.~\ref{fig:esr1}). At 5~K, more than 20 lines are visible in the spectrum. This complex structure is typical in low-temperature ESR spectra of molecular magnets and is related to the hyperfine coupling.\cite{cage2003,nellutla2005}
The complexity of the signal and the presence of multiple Cu sites and anisotropy parameters prevent us from a detailed analysis of the low-temperature powder spectra. One would need much higher frequencies and, preferably, single crystals of \cunipa, in order to discriminate the multiple resonances that are visible below 40~K. So we restrict ourselves to the analysis of the data above 40~K. 
\begin{figure}[!t]
\includegraphics{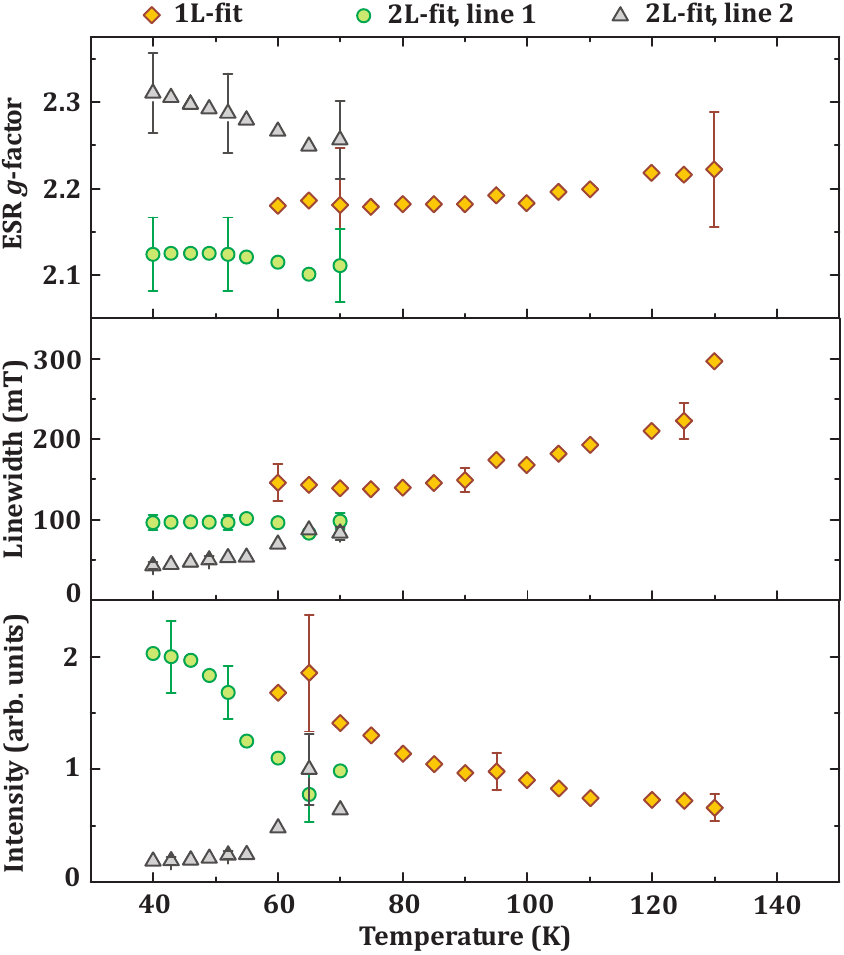}
\caption{\label{fig:esr2}
(Color online) Temperature dependence of the ESR signals: $g$-factor (top), linewidth (middle), intensity (bottom). The data at higher temperatures (diamonds) were obtained by fitting the spectra with one Lorentzian line (1L-Fit). Open circles and triangles show the data from fitting the spectra with two Lorentzian lines (2L-Fit). Representative error bars are shown for several data points, only. All intensities are scaled to the intensity of line 2 at 65~K.
}
\end{figure}

Fig.~\ref{fig:esr2} shows the temperature evolution of the ESR $g$-factor, linewidth, and line intensities. Above 70~K, the observed powder-averaged values of $g\simeq 2.2$ are typical for Cu$^{2+}$ in the planar oxygen environment.\cite{choi2006,*choi2008,cage2003,nellutla2005} Our microscopic insight into the energy spectrum of the Cu$_5$ molecule (Sec.~\ref{sec:ED}) suggests that the formation of two lines below 70~K is related to two lowest $S\!=\!\frac12$ energy levels, which are separated by $\Delta\simeq 68$~K. Indeed, lines 1 and 2 show different temperature evolution. While the intensity of line 2 is reduced upon cooling, the intensity of line 1 is growing. In Sec.~\ref{sec:ED} below, we provide the explicit dependence of the effective $g$-tensors for the GS and the first excited doublet (in terms of the individual $g$-tensors of the three inequivalent Cu sites) in order to demonstrate the origin of their difference.

\subsection{$^{1}$H NMR}
\label{sec:nmr}
\subsubsection{Linewidth}\label{sec:nmrLineWidth}
\begin{figure}[!t]
\includegraphics{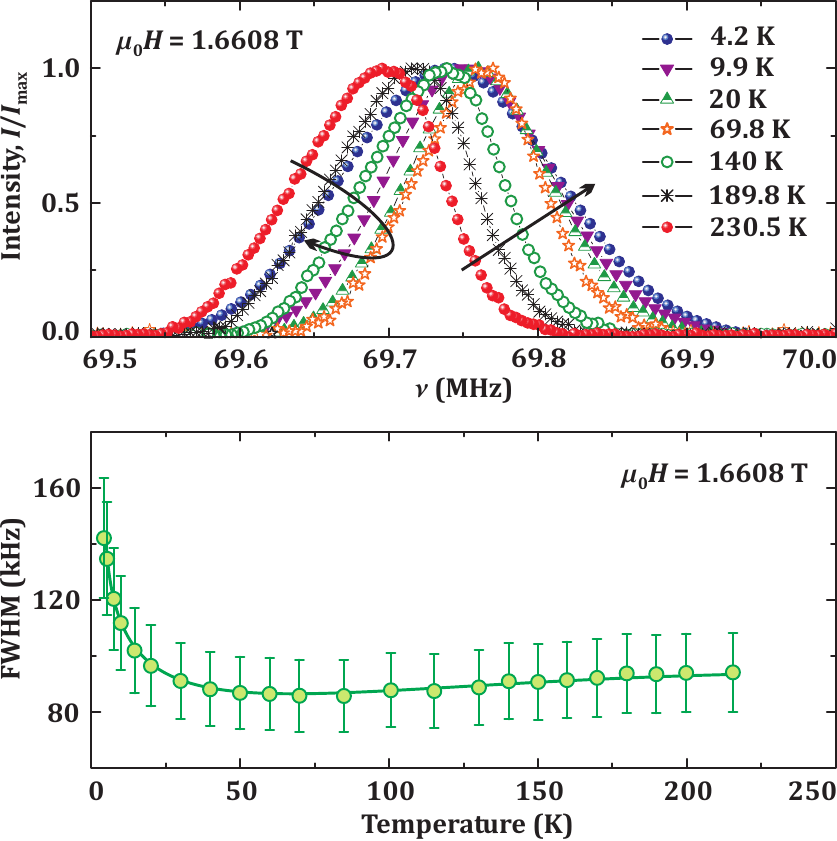}
\caption{\label{fig:linewidth} (Color online) Top panel: Fourier-transform $^{1}$H NMR spectra measured at different temperatures. Arrows show the shifts of the left and right shoulders upon cooling. Note that while the right shoulder is moving toward higher frequencies, the left shoulder shows a non-monotonic behavior. Bottom panel: Full width at half-maximum (FWHM) of $^{1}$H NMR spectra plotted as a function of temperature ($T$) measured in an external field $\mu_0H=1.6608$~T.
}
\end{figure}

Considering the nuclear spin $I=\frac12$ of the $^{1}$H nucleus, one expects a single spectral line for each of the 17 nonequivalent proton sites (see Table~\ref{tab:structuraldata} of App.~\ref{app:B}). However, these lines merge into a single broad line with a nearly Gaussian line shape over the whole measured temperature range (Fig.~\ref{fig:linewidth}, top). The linewidth at 215~K is 95~kHz and almost comparable with the linewidth reported for other molecular magnets (see, e.g., Ref.~\onlinecite{khuntia2010}). The line position shifts weakly with temperature. The bottom panel of Fig.~\ref{fig:linewidth} shows the temperature evolution of the full width at half-maximum (FWHM) of the $^1$H NMR line measured in an applied field of $\mu_0H=1.6608$~T. It is almost temperature-independent at high temperatures and increases progressively upon cooling below about 30~K.

The shape and width of the $^{1}$H NMR spectra are governed by two main interactions: the nuclear-nuclear dipolar interactions and the hyperfine couplings between the proton and unpaired electrons at the Cu sites. Therefore, we can write FWHM as:\cite{slichter1996,abragamNMR,belesi2009}  
\begin{equation}
\text{FWHM} \propto \sqrt{\langle \Delta \nu^2\rangle_d + \langle \Delta \nu^2\rangle_m},
\label{FWH}
\end{equation}
where the broadening due to the nuclear-nuclear dipolar interactions ($\langle\Delta\nu^2\rangle_d$) is temperature-independent, 
while the broadening due to the hyperfine couplings ($\langle \Delta \nu^2\rangle_m$) scales with the local susceptibilities.   
For each given proton site $p$, 
\begin{equation}
\frac{\sqrt{\langle \Delta \nu^2\rangle_{m}^{(p)}}}{H}\simeq \sum_j A_{j}^{(p)}\chi_j ,
\label{moment}
\end{equation}
where $A_{j}^{(p)}$ is the dipolar coupling constant between the proton and the Cu$^{2+}$ ions at site $j$, and $\chi_j$ is the local susceptibility. We should note here that, in general, the temperature dependence of $\chi_j$ is different from that of the bulk susceptibility $\chi$ (Fig.~\ref{fig:chi}), especially at higher temperatures, and they are also different from each other (see Fig.~\ref{fig:localMomentsvsT}). The sharp increase in FWHM at low temperatures can be ascribed to the corresponding low-temperature increase in $\chi_j$. 

The shallow minimum  in FWHM observed around 70~K is paralleled by the peculiar evolution of the spectral line (Fig.~\ref{fig:linewidth}, top). On cooling down, the right side of the line shifts weakly but monotonously to the right, following the central position of the line. The left side, on the other hand, shows a non-monotonic behavior, shifting backwards for the data at 20 K and below. The origin of this feature (and the minimum in FWHM) can in principle be traced back to the temperature dependence of the local Cu$^{2+}$ moments, and indeed such a non-monotonic behavior is shown by the moments on the Cu2 sites. As we discuss in detail below (Sec.~\ref{sec:ED} and Fig.~\ref{fig:localMomentsvsT}), at $T\!\ll\!\Delta\simeq 68$ K, the moments of the two Cu2 sites are antiparallel to the field, owing to the large negative exchange field exerted by the neighboring Cu1 and Cu3 sites. These exchange fields are balanced by entropy at some characteristic temperature $T^\ast$, at which the Cu2 moments turn positive. Our calculations based on the actual exchange couplings give $T^*\simeq 38$ K (see Fig.~\ref{fig:localMomentsvsT} below). At even higher temperatures, the Cu2 moments attain their paramagnetic Curie-like behavior of isolated spins. The contribution to the second moment from the Cu2 sites is then expected to decrease down to zero and then increase again as we cool down, starting from the high-temperature paramagnetic to the low-temperature ``ferrimagnetic'' $S=\frac12$ state of \cunipa.

\subsubsection{Wipe-out effect}
\begin{figure}
\includegraphics{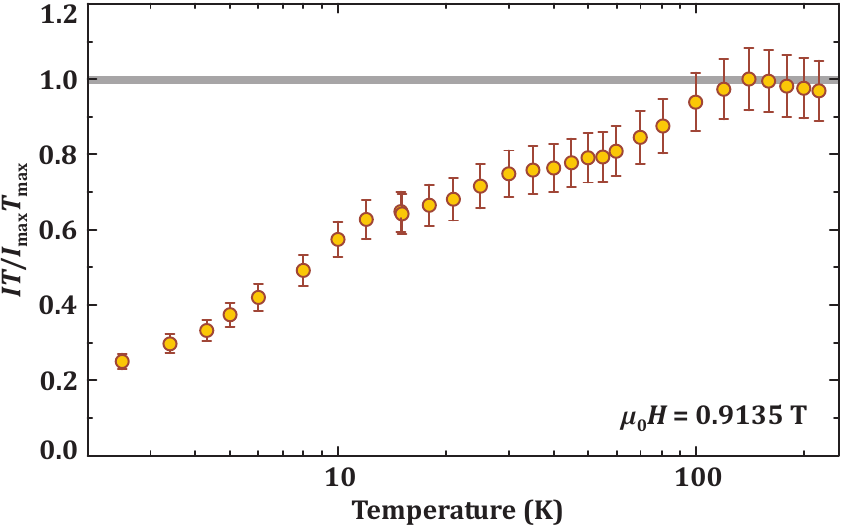}
\caption{\label{II} (Color online) Normalized integrated intensity ($IT/I_{\max}T_{\max}$) measured at $\mu_0H=0.9135~$T. 
The strong drop in the intensity below 100~K is the wipe-out effect. The line shows the maximum intensity ($I_{\max}T_{\max}$) 	attained at high temperatures.}
\end{figure}

The wipe-out effect is common in molecular nanomagnets,\cite{belesi2005, borsa2006,khuntia2011,amiri2010} 
and refers to the gradual loss in the NMR signal intensity below a characteristic temperature. 
Here, we have measured the temperature dependence of $M_{xy}(0)T$, where $M_{xy}(0)$ is the transverse magnetization at time $t\!=\!0$, obtained by the extrapolation of the $M_{xy}(t)$ recovery curve to $t\!=\!0$. The integrated intensity $M_{xy}(0)T$ is proportional to the total number of protons resonating at the irradiated frequency. 
The normalized integrated intensity of the NMR signal in \cunipa\ as a function of temperature is shown in Fig.~\ref{II}. 
We find that the loss of the NMR signal begins below 150~K and becomes more pronounced at low temperatures. The onset of the NMR signal loss around 150~K coincides with the regime where $1/T_1$ starts increasing towards the maximum (see Fig.~\ref{t1} below).

The loss in the NMR signal is related to the slowing down of the electron spin dynamics which, as we explain below, 
is also responsible for the enhancement in $1/T_1$. On decreasing $T$, a progressively larger fraction of protons close to the magnetic Cu$^{2+}$ ions attain spin-spin relaxation times ($T_{2}$) shorter than the dead time ($\tau_{d}$) of the spectrometer, hence the signal from these protons cannot be detected. At low temperatures, only the protons that are very far from the magnetic ions contribute to the signal intensity.  For further details, see Refs. \onlinecite{belesi2005, borsa2006}. 

\subsubsection{Nuclear spin-lattice relaxation rate $1/T_1$}
\begin{figure}[!t]
\includegraphics{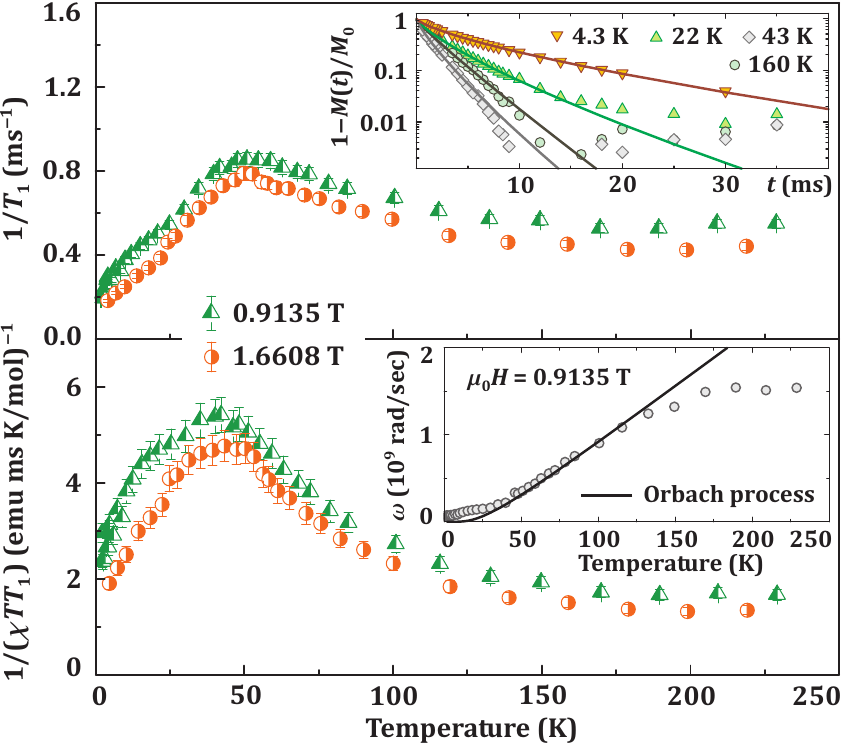}
\caption{\label{t1} (Color online) Top panel: Spin-lattice relaxation rate $1/T_1$ vs. $T$ measured at two different values of external field, 0.9135~T and 1.6608~T. The inset shows the longitudinal recovery curves at four representative temperatures; solid lines are the fits using Eq.~\eqref{exp}. Bottom panel: $1/(\chi T_1 T)$ vs. $T$ for two different magnetic fields. The inset shows the temperature-dependent $\omega_c$ (decay rate of the total moment $S_z$) extracted from the $1/T_1$ data; the solid line is the fit using Eq.~\eqref{Orbach}.}
\end{figure}

The spin-lattice relaxation rate $1/T_{1}$ for Cu$_5$-NIPA was measured at two different applied fields. 
The recovery of the longitudinal nuclear magnetization after a saturation pulse was fitted well by the stretched exponential function (top panel of Fig.~\ref{t1}, inset)
\begin{equation}
1-\frac{M(t)}{M_{0}}\simeq A' e^{-(t/T_{1})^\beta},
\label{exp}
\end{equation}
where $M(t)$ is the nuclear magnetization at a time $t$ after the saturation pulse, and $M_{0}$ is the equilibrium magnetization. 
The value of the exponent $\beta$ for both fields was found to decrease slowly from 0.95 to 0.7 upon cooling. The deviation of $\beta$ from unity reflects the distribution of relaxation rates according to different hyperfine couplings between the $^{1}$H nuclei and Cu$^{2+}$ ions.\cite{borsa2006,khuntia2009,lowe1968} Accordingly, the spin-lattice relaxation rate $1/T_{1}$ obtained by fitting the recovery with Eq.~\eqref{exp} provides a value averaged over the whole distribution. 

The $T$ dependence of $1/T_{1}$ is presented in the top panel of Fig.~\ref{t1} for two values of the external field. 
At high temperatures ($T\gtrsim 150$~K), $1/T_{1}$ is almost $T$-independent. 
This behavior is typical for uncorrelated paramagnetic moments fluctuating fast and at random.\cite{moriya1956,belesi2007} 
At lower temperatures, $1/T_{1}$ increases and passes through a maximum at $T\simeq 40$~K. 
Such a characteristic enhancement has been found in numerous AFM homometallic\cite{baek2004} 
and heterometallic\cite{amiri2010} rings built of spins $s\!>\!\frac12$, but in spin-$\frac12$ systems it is very rare. The only example known to us is the Cu$_6$ magnet with a high-spin $S=3$ GS.\cite{lascialfari1998,carretta2006} Systems with predominantly AFM couplings and low-spin GS (e.g., V$_{12}$ having an $S=0$ GS, Ref.~\onlinecite{procissi2004}) do not show such a feature in $1/T_1$.

In homometallic rings, the maximum in $1/T_{1}$ essentially signals the slowing down of phonon-driven spin fluctuations.\cite{borsa2006,baek2004,santini2005,ioannisPRB2007,ioannisPRB2009} The equivalence between the spins (by virtue of the nearly perfect rotation symmetry of the ring) allows to express $1/T_1$ in terms of the spectral density of the total magnetic moment $S_z$ of the molecule.\cite{santini2005} As shown numerically\cite{santini2005} and by a microscopic theory,\cite{ioannisPRB2007,ioannisPRB2009} the phonon-driven decay of $S_z$ proceeds independently from the remaining observables of the problem, which in turn leads to a single Lorentzian form for $1/T_1$:
\begin{equation}
\frac{1}{T_{1}} \simeq A\chi T \frac{\omega_{c}(T)}{\omega_{c}^{2}(T)+\omega_{L}^{2}},
\label{bpp}
\end{equation}
where $A$ is the average square of the transverse hyperfine field, $\omega_{L}$ is the nuclear Larmor frequency, 
and $\omega_{c}= \frac{1}{\tau (T)}$ is the decay rate of the total moment $S_z$. The enhancement in the spin-lattice relaxation rate $1/T_1$ takes place when $\omega_c$ approaches the order of magnitude of $\omega_{L}$.

There are two qualitative differences between \cunipa\ and homometallic rings that should be emphasized, though. 
First, Cu$_5$-NIPA features a very sparse excitation spectrum (Sec.~\ref{sec:ED}), which means that there are very few spin-phonon channels available for relaxation. According to Ref. \onlinecite{ioannisPRB2009}, this also means that here $\omega_c$ is not expected to show the strong power-law $\sim T^3$ or $T^4$ behavior. Instead, one expects a much weaker $T$ dependence, with longer relaxational times in a wide temperature range.

The second difference is the fact that the Cu$_5$ molecule comprises three inequivalent Cu sites with different local magnetizations, and so the above single-Lorentzian formula for $1/T_1$ is not valid any longer, but instead a multi-Lorentzian form should be expected on general grounds, as in the case of the heterometallic Cr$_7$Ni ring.\cite{bianchi2010}

Given the large spin gap, $\Delta\simeq 68$~K, one could still argue in favor of the single-Lorentzian formula of Eq.~\eqref{bpp} 
over a wide low-$T$ range, since the Wigner-Eckart theorem allows to replace individual spin operators with the total spin operators times a constant. However, such a treatment would only capture the resonant spin-phonon transitions between the two Zeeman-split levels of the GS, but these transitions are prohibited by Kramer's theorem. Instead, for temperatures well below the spin gap $\Delta\simeq 68$ K, the relaxation dynamics of the spins must be controlled by inter-multiplet Orbach transitions\cite{abragamEPR} between the GS doublet and the lowest magnetic excitation with the energy $\Delta$ (see below). Again, this is quite similar to Cr$_7$Ni, with the only difference that the spin gap is much smaller there ($\simeq 14$ K).\cite{bianchi2010}

With the above remarks in mind, we may still use the above single-Lorentzian formula for $1/T_1$ to extract the temperature dependence of $\omega_c$, but the latter is now a representative measure of the relaxation dynamics (as probed by NMR), and not necessarily the relaxation rate of the total moment. For simplicity, we have plotted $1/\chi T_1T$ vs. $T$ in the bottom panel of Fig.~\ref{t1} that clearly shows a broad maximum reflecting the slowing down of fluctuating moments. Following Eq.~\eqref{bpp} and the procedure outlined in Ref.~\onlinecite{baek2004}, we extracted $\omega_c$ from the $1/\chi T_1T$ data at $\mu_0H=0.9135$~T. The resulting $T$ dependence of $\omega_c$ (shown in the lower inset of Fig.~\ref{t1}) confirms that it is much weaker compared to the strong power-law dependence found in homometallic rings.\cite{baek2004} 
In particular, there is a wide low-$T$ range over which $\omega_c$ follows the inter-multiplet Orbach relaxation processes mentioned above. Following Ref.~\onlinecite{abragamEPR},
\begin{equation}
\omega_c \simeq \frac{3 \lambda^2 \Delta^3}{2\pi\hbar^4 \rho_m v^5}\times\frac{1}{e^{\Delta/T}-1},
\label{Orbach}
\end{equation}
where $\rho_m=2043$~kg/m$^3$ is the mass density,\cite{liu2011} $v$ is the sound velocity in \cunipa\ 
(typical value $c\sim1500$ m/sec in nanomagnets\cite{garanin2008}), and $\lambda$ stands for the spin-phonon coupling energy parameter, which is related e.g. to the fluctuating portion of the Dzyalozinskii-Moriya interactions present in this system.  Note that we neglect the Zeeman splitting contribution to the resonance energy, because it is negligible compared to $\Delta$. 
By fitting the $\omega_c$ data with Eq.~\eqref{Orbach}, we obtain $\lambda/k_B \simeq 3.9 (v/c)^{\frac{5}{2}}$~K.

A more complete quantitative understanding of $1/T_1$ (e.g. the high-temperature behavior and field dependence) 
must take into account the multi-exponential behavior of the relaxation, discussed above, but also the presence of the 
wipe-out effect discussed above, see e.g. Ref. \onlinecite{bianchi2010}. 

\section{Theory}\label{sec:theory}
\subsection{Microscopic magnetic model}\label{sec:model}
\begin{figure}
\includegraphics{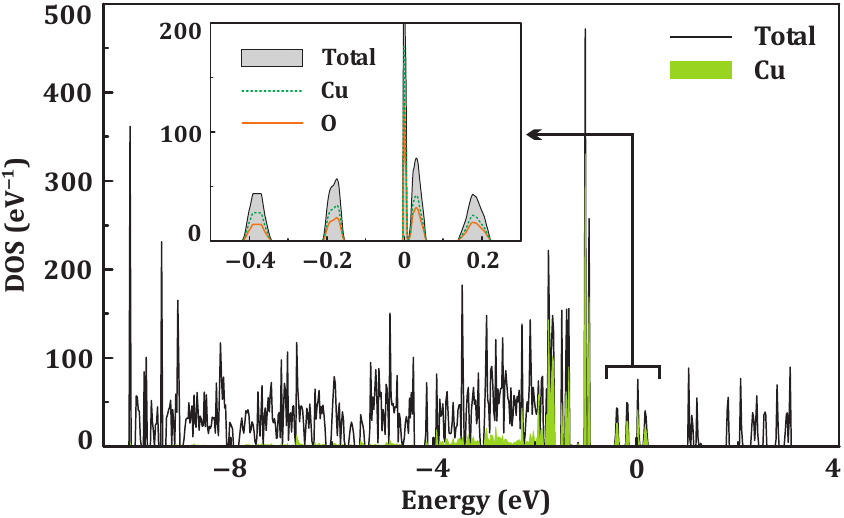}
\caption{\label{fig:dos}
(Color online) LDA density of states for \cunipa. The shading shows the contribution of Cu orbitals. The inset magnifies five bands at the Fermi level ($E=0$), with the atomic contributions denoted by the dashed (Cu) and solid (O) lines.
}
\end{figure}
To determine individual exchange parameters in the Cu$_5$ molecule, we calculate the electronic structure of the \cunipa\ compound. This procedure requires reliable crystallographic information, including precise positions of all atoms in the unit cell. However, the available structural data\cite{liu2011,zhao2011} are obtained from x-ray diffraction that has only limited sensitivity to the positions of hydrogen atoms. Therefore, we used the literature data as a starting model and optimized the hydrogen positions, whereas all other atoms were kept fixed. The equilibrium positions of hydrogen are listed in Table~\ref{tab:structuraldata} of App.~\ref{app:B}. The relaxed structure is 18.6~eV/f.u. lower in energy than the starting model taken from the literature. This energy reduction should be ascribed to the elongation of O--H distances that are unrealistically short (about 0.8~\r A) in the experimental structural data.\cite{liu2011}

\begin{figure}
\includegraphics{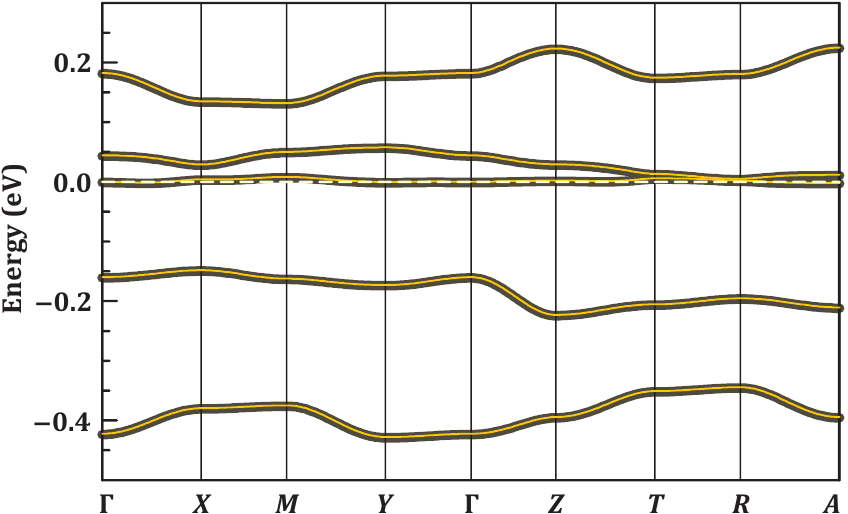}
\caption{\label{fig:bands}
(Color online) LDA band structure of \cunipa\ (thin light lines) and the fit with the tight-binding model (thick dark lines). The Fermi level is at zero energy (white dashed line). The notation of $k$ points is as follows: $\Gamma(0,0,0)$, $X(\frac12,0,0)$, $M(\frac12,\frac12,0)$, $Y(0,\frac12,0)$, $Z(0,0,\frac12)$, $T(\frac12,0,\frac12)$, $R(\frac12,\frac12,\frac12)$, and $A(0,\frac12,\frac12)$, where the coordinates are given in units of the reciprocal lattice parameters.
}
\end{figure}

The LDA energy spectrum of \cunipa\ (Fig.~\ref{fig:dos}) comprises narrow lines that represent molecular orbitals of the NIPA molecules and the Cu$_5$(OH)$_2$ unit. The states at the Fermi level belong to the narrow Cu $3d$ bands with a sizable admixture of O $2p$. The local environment of Cu atoms resembles the conventional CuO$_4$ plaquette units (four-fold coordination, see Fig.~\ref{fig:structure}, left). Therefore, the highest-lying Cu $3d$ bands have predominantly $x^2-y^2$ origin, following the crystal-field levels of Cu$^{2+}$ with $x$ and $y$ axes lying in the plane of the CuO$_4$ plaquette. The five Cu atoms of the Cu$_5$ molecule (one molecule per unit cell) give rise to five bands, with the middle band crossing the Fermi level (Figs.~\ref{fig:dos} and~\ref{fig:bands}). This spurious metallicity is due to the strong underestimate of electronic correlations in LDA. DFT+$U$ calculations reveal the robust insulating behavior with a band gap of about 2.2~eV in reasonable agreement with the light-blue color of the sample.

\begin{table}
\caption{\label{tab:couplings}
Interatomic distances $d$ (in~\r A), Cu--O--Cu bridging angles $\varphi$ (in~deg), hopping parameters $t_i$ (in~meV), and exchange integrals $J_i$ (in~K) in \cunipa. The $J_i$ values are obtained from DFT+$U$ calculations. The AFM contributions $J_i^{\AFM}$ are evaluated as $4t_i^2/U_{\eff}$, where $U_{\eff}$ is the effective on-site Coulomb repulsion. The FM contributions are $J_i^{\FM}=J_i-J_i^{\AFM}$.
}
\begin{ruledtabular}
\begin{tabular}{ccccrrr}
          & $d_{\text{Cu--Cu}}$ & $\varphi_{\text{Cu--O--Cu}}$ & $t_i$    & $J_i^{\AFM}$ & $J_i^{\FM}$ & $J_i$ \\
 $J_{13}$ &    3.20             &             107.9            & $-0.054$ &    34        &   $-89$     & $-55$ \\
 $J_{12}$ &    3.33             &             114.5            & $-0.128$ &    191       &  $-134$     & 57    \\    
 $J_{23}$ &    3.51             &             125.7            & $-0.161$ &    302       &   $-34$     & 268   \\
\end{tabular}
\end{ruledtabular}
\end{table}

To evaluate the magnetic couplings, we fit the five Cu bands with a tight-binding model (Fig.~\ref{fig:bands}) and extract the relevant hopping parameters $t_i$ using Wannier functions (WFs) based on the Cu $d_{x^2-y^2}$ orbital character.\cite{wannier} The $t_i$'s are further introduced into an effective Hubbard model with the on-site Coulomb repulsion $U_{\eff}$. As the conditions of the half-filling and strong correlations ($t_i\ll U_{\eff}$) are fulfilled, the lowest-lying excitations can be described by a Heisenberg model with the AFM exchange $J_i^{\AFM}=4t_i^2/U_{\eff}$. The values of $t_i$ and $J_i^{\AFM}$ obtained from the LDA band structure are listed in Table~\ref{tab:couplings} and provide an overview of the magnetic couplings in \cunipa. Considering only nearest-neighbor interactions, we find a strong coupling between Cu2 and Cu3, a somewhat weaker coupling between Cu1 and Cu2, and a relatively weak coupling between Cu1 and Cu3. Long-range couplings are several times smaller than $J_{13}^{\AFM}$. The long-range exchanges in the Cu$_5$ molecule are within 10~K, whereas the interactions between the molecules are below 3~K (note the experimental estimate of 0.5~K in Sec.~\ref{sec:thermo}). As the long-range couplings are much weaker than $J_{12}$, $J_{13}$, and $J_{23}$, we further restrict ourselves to the minimum microscopic model that comprises three nearest-neighbor interactions, only.

\begin{figure}
\includegraphics{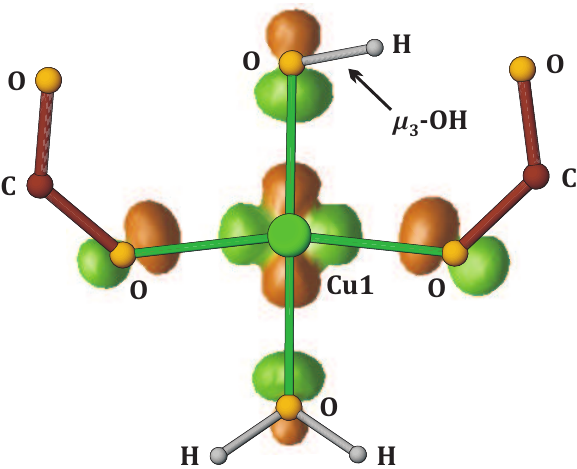}
\caption{\label{fig:wannier}
(Color online) Cu $d_{x^2-y^2}$-based Wannier function for the Cu1 site in \cunipa. Note that the ligands (COO$^-$, $\mu_3$-OH$^-$, H$_2$O) affect the positions of oxygen atoms and their orbitals. However, only the orbitals of four oxygen atoms  contribute to the Wannier function.
}
\end{figure}
Possible FM contributions to the short-range couplings require that the evaluation of $J_i^{\AFM}$ is supplied by an independent estimate of total exchange couplings $J_i$. In a so-called supercell approach, total energies of collinear spin configurations are mapped onto the Heisenberg model to yield the $J_i$ values, as listed in Table~\ref{tab:couplings}. All nearest-neighbor interactions have sizable FM components $J_i^{\FM}=J_i-J_i^{\AFM}$ that reduce $J_{12}$ and $J_{23}$, whereas $J_{13}$ eventually becomes ferromagnetic. This way, the triangular units in the Cu$_5$ molecule feature a combination of FM interaction $J_{13}$ and AFM interactions $J_{12}$ and $J_{23}$ (Fig.~\ref{fig:structure}, left). 

The microscopic origin of the magnetic couplings can be understood from the analysis of Cu-based Wannier functions (Fig.~\ref{fig:wannier}). Each WF features the Cu $d_{x^2-y^2}$ orbitals together with the $2p$ orbitals of the surrounding oxygen atoms. These oxygen atoms have different chemical environment and belong to one of the ligands: $\mu_3$-OH$^-$, COO$^-$, and H$_2$O. While the ligands affect the positions of oxygen atoms (note the slight downward displacement of the O atoms in COO$^-$ groups) and modify the ``shape'' of the $p$ orbitals (note the oxygen atom belonging to the H$_2$O molecule), they do not provide any sizable long-range contributions to the WFs. The short-range nature of the WFs underlies the diminutively small long-range exchange in \cunipa. Further microscopic aspects of the magnetic interactions in \cunipa\ are discussed in Sec.~\ref{sec:discussion}. 

\begin{figure}
\includegraphics{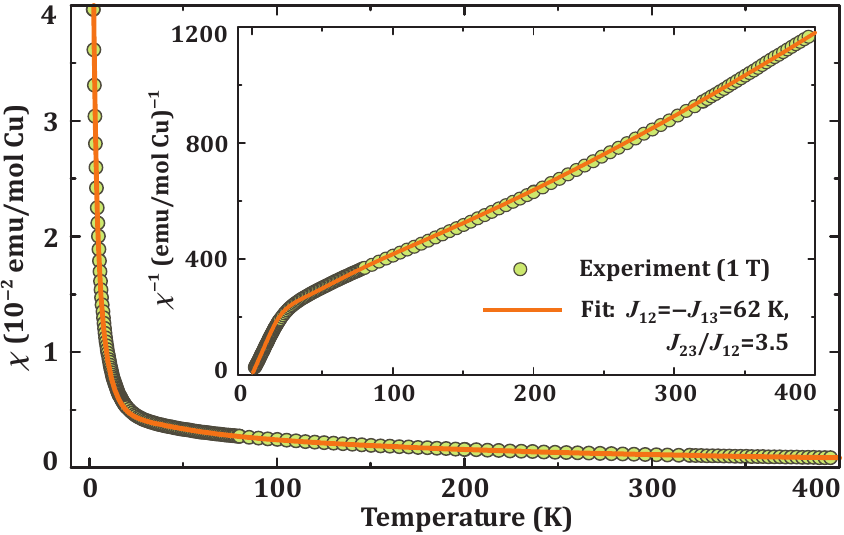}
\caption{\label{fig:chi-fit}
(Color online) Fit of the magnetic susceptibility ($\chi$) and inverse magnetic susceptibility ($\chi^{-1}$, inset) with the exact-diagonalization result for the Cu$_5$ pentamer ($J_{12}=-J_{13}=65$~K and $J_{23}/J_{12}=3.5$).
}
\end{figure}

Our microscopic model can be directly compared to the experimental data. Using full diagonalization for the Cu$_5$ molecule, we calculate thermodynamic properties and refine the $J_i$ parameters as to match the experiment. This way, we fit the experimental magnetic susceptibility curve with $J_{12}=-J_{13}=62$~K and $J_{23}=217$~K (so that $J_{12}/J_{23}\simeq\frac27$), as well as $g=2.22$ and the temperature-independent contribution $\chi_0=-1.7\times 10^{-4}$~emu/mol~Cu (Fig.~\ref{fig:chi-fit}). The experimental estimates of $J_i$ are in excellent agreement with the DFT results (Table~\ref{tab:couplings}), whereas the $g$-value matches the high-temperature $g\simeq 2.2$ from ESR (Sec.~\ref{sec:esr}). Regarding the sizable $\chi_0$, it may originate from the diamagnetic gelatin capsule that was used in the susceptibility measurement.

\begin{figure}
\includegraphics{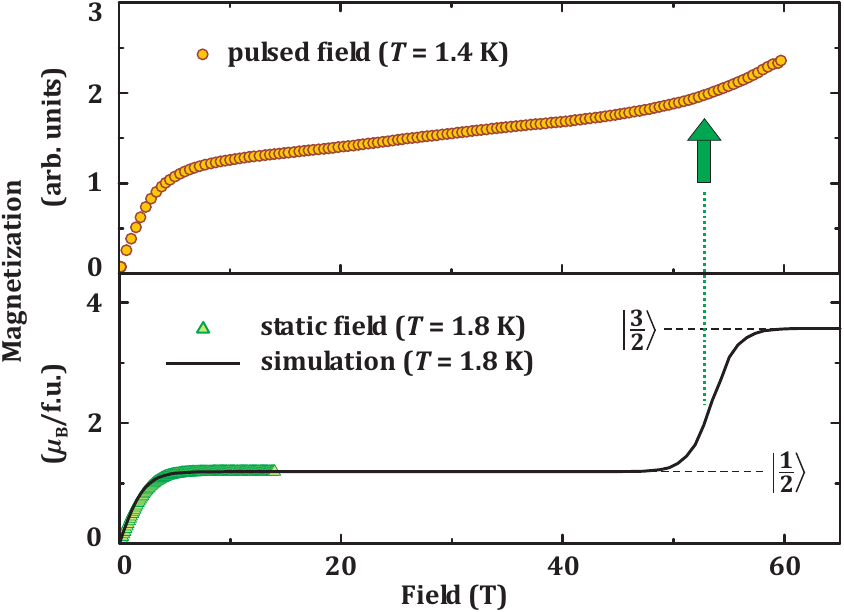}
\caption{\label{fig:mvsh-fit}
(Color online) Upper panel: magnetization of \cunipa\ measured in pulsed magnetic field (in~arb.~units). Bottom panel: magnetization measured in static field up to 14~T and the simulated curve for $J_{12}=-J_{13}=65$~K and $J_{23}/J_{12}=3.5$ (both in units of $\mu_B$/f.u.) The arrow denotes the bend in the pulsed-field curve that matches the transition between the $\onehalf$ and $\threehalf$ states of the Cu$_5$ molecule.
}
\end{figure}
The combination of the FM coupling $J_{13}$ and AFM couplings $J_{12}$ and $J_{23}$ implies that the spin triangles in Cu$_5$ are non-frustrated. In a classical picture, all three couplings are satisfied in a configuration with Cu3 and Cu1 spins pointing up and Cu2 spins pointing down, thus leading to the total moment of $\frac12$ per molecule. Indeed, at low temperatures \cunipa\ shows paramagnetic behavior with the magnetic moment of $\frac12$ per molecule ($\onehalf$ state). In high magnetic fields, the $\threehalf$ and, eventually, $\fivehalf$ states will be stabilized. The formation of states with the higher magnetic moment can be seen from the simulated magnetization curve of \cunipa. The $\onehalf$ state persists up to about 50~T, where the magnetization increases abruptly until the $\threehalf$ state is reached. 

We endeavored to verify this prediction with magnetization measurements in pulsed fields. Unfortunately, the results are strongly affected by dynamic effects. The typical duration of the pulse (about 20~$\mu$s) is insufficient to change the magnetic moment of the molecule. Therefore, all features are blurred and, moreover, the curve shows a non-zero slope above 5~T (Fig.~\ref{fig:mvsh-fit}, top), in contrast to the flat plateau that is observed in static field (Fig.~\ref{fig:mvsh-fit}, bottom). The apparent mismatch between the pulsed-field and static-field data also prevents us from scaling the high-field curve. Nevertheless, the bend observed at $50-55$~T is likely the signature of the $\onehalf\longrightarrow\threehalf$ transition and confirms our expectations.

\subsection{Exact Diagonalization}\label{sec:ED}
We are now ready to evaluate the magnetic energy spectrum of each Cu$_5$ molecule and deduce its basic properties, according to the DFT values of the exchange couplings $J_i$. Using a site labeling convention in accord with the site symmetries, 
the spin Hamiltonian (disregarding anisotropies) reads: 
\be\label{eq:Ham}
\mc{H}\!=\! J_{23} \vec{S}_3\!\cdot\!\vec{S}_{22'} 
+ J_{12} \left(\vec{S}_1\!\cdot\!\vec{S}_2+\vec{S}_{1'}\!\cdot\!\vec{S}_{2'} - \vec{S}_3\!\cdot\!\vec{S}_{11'} \right),
\ee 
where $\vec{S}_{22'}\equiv \vec{S}_2+\vec{S}_{2'}$, $\vec{S}_{11'}\equiv \vec{S}_1+\vec{S}_{1'}$, 
and we have used the relation $J_{13}=-J_{12}$. 
In addition to the full SU(2) spin rotation group, this Hamiltonian is also invariant under 
the inversion operation (in real space) through the central site $\vec{S}_3$, which maps 
$\vec{S}_3\!\leftrightarrow\!\vec{S}_3$, $\vec{S}_1\!\leftrightarrow\!\vec{S}_{1'}$, and $\vec{S}_2\!\leftrightarrow\!\vec{S}_{2'}$. 
So all eigenstates can be characterized by the total spin $S$, its projection $M$ along 
some quantization axis $\vec{z}$, and the parity $p$ (even or odd) under inversion. 

{\it Global spectral structure.}
A straightforward numerical diagonalization of $\mc{H}$ yields the energy spectrum given in Table \ref{tab:ExactSpectrum}.  
For each level we also provide the good quantum numbers $S$ and $p$, as well as the expectation values of 
the square of the composite spins $\vec{S}_{22'}$, $\vec{S}_{322'}\equiv\vec{S}_3\!+\!\vec{S}_{22'}$, and $\vec{S}_{11'}$.
The essential finding is that the GS is a total spin $S\!=\!\frac12$ doublet and the lowest excitation (also a $S\!=\!\frac12$ doublet) has a very high energy gap $\Delta\!=\!0.313 J_{23}\!\simeq\! 68$~K. Thus the Cu$_5$ molecule behaves as a rigid $S\!=\!\frac12$ entity for $T\ll\Delta$, confirming the previous experimental picture from magnetization measurements (Sec.~\ref{sec:chi}). 

Looking at Table~\ref{tab:ExactSpectrum}, we also find that the spectrum is organized in three compact groups of states,  
which are separated by horizontal lines. All states belonging to a given group show a very similar value of $\langle\vec{S}_{322'}^2\rangle$, 
suggesting that the trimer of $\vec{S}_3$, $\vec{S}_2$, and $\vec{S}_{2'}$ plays a special role in the physics of Cu$_5$-NIPA.
Moreover, the expectation values of $\vec{S}_{22'}^2$, $\vec{S}_{322'}^2$, and $\vec{S}_{11'}^2$  
are for all states very close to the formula $S(S+1)$ for some integer or half-integer $S$, suggesting that  
these composite spins are almost conserved quantities. Below, we shall demonstrate that all these spectral features as well as the GS properties can be physically understood quite naturally on the basis of a strong coupling expansion around the limit of $J_{12}\!=\!|J_{13}|\!=\!0$.

From the spectrum, we can also obtain the level-crossing fields at which the GS of the molecule changes from $S=\frac12$ to $S=\frac32$ and from $S=\frac32$ to $S=\frac52$. We find: 
\be
g\mu_B H_{c1}/k_B \simeq 0.40 J_{23} , ~~~
g\mu_B H_{c2}/k_B \simeq 1.43 J_{23} . 
\ee
With $J_{23}=217$ K and the average $g\simeq 2.38$ (see previously), these numbers yield $H_{c1}\simeq 54$ T (compare to Fig.~\ref{fig:mvsh-fit}) and $H_{c2}\simeq 194$ T.   

\begin{table}[!t]
\caption{\label{tabExactSectrum}
The exact spectrum of the spin Hamiltonian $\mc{H}$ of Eq.~(\ref{eq:Ham}) at $y=J_{12}/J_{23}=\frac27$.  
For each state we also provide its total spin $S$, its parity $p$ under the real-space inversion through the central site $\vec{S}_3$, as well as the expectation values of $\vec{S}_{22'}^2$, $\vec{S}_{22'}^2$, and $\vec{S}_{22'}^2$. The horizontal lines differentiate the three unperturbed manifolds of the strong coupling limit $y\!=\!0$ (see text and Fig.~\ref{fig:SpectrumMomentsCorrsvsy}). 
}
\begin{ruledtabular}\label{tab:ExactSpectrum}
\begin{tabular}{ccrrrr}
total S  & parity $p$ & total $E/J_{23}$  & $\langle\vec{S}_{22'}^2\rangle$ & $\langle\vec{S}_{322'}^2\rangle$ 
& $\langle\vec{S}_{11'}^2\rangle$ \\
\hline
$\frac{1}{2}$ & even &  $-1.3289$   & 1.969 & 0.795 & 1.969 \\
$\frac{1}{2}$ & odd & $-1.0157$  & 1.976 & 0.75 & 0.024 \\
$\frac{3}{2}$ & even & $-0.9286$  & 2 & 0.893 & 2 \\
\hline
$\frac{3}{2}$ & odd & $-0.1732$  & 0.086 & 0.879 & 1.914 \\
$\frac{1}{2}$ & even & $-0.0395$  & 0.222 & 1.011 & 0.222\\
$\frac{1}{2}$ & odd &  0.3014 & 0.024& 0.75 & 1.976 \\
\hline
$\frac{5}{2}$ & even & 0.5 & 2 & 3.75 & 2\\
$\frac{3}{2}$ & odd & 0.5303 & 1.914& 3.621& 0.086 \\
$\frac{3}{2}$ & even & 0.5714 & 2 & 3.607 & 2\\
$\frac{1}{2}$ & even & 0.5827 & 1.809& 3.444& 1.8087
\end{tabular}
\end{ruledtabular}
\end{table} 
  
{\it GS properties.}
To probe the nature of the GS, we calculate several experimentally relevant GS expectation values. We first consider how the total $S\!=\!\frac12$ moment of the GS (at low $T$, this moment can be attained by a small applied field) is distributed among the five Cu sites. At $T\!\ll\!\Delta$, we find 
\be\label{eq:ExactMoments}
\langle S^z_i\rangle = \langle S^z_i\rangle_0 \times \tanh\left(\frac{g\mu_B H}{2k_BT}\right), 
\ee
where $\langle S^z_i\rangle_0$ are the GS expectation values (in an infinitesimal field): 
\be\label{eqn:GSlocalmoments}
\begin{array}{c}
\langle S_3^z\rangle_0\!=\!0.1416,\quad \langle S_2^z\rangle_0=\langle S_{2'}^z \rangle_0\!=\!-0.1415, \\
\langle S_1^z \rangle_0=\langle S_{1'}^z \rangle_0\!=\!0.3207. 
\end{array}
\ee
The signs of the above numbers are in agreement with the classical ferrimagnetic GS discussed above (Fig.~\ref{fig:structure}, right), whereby the spins $\vec{S}_3$, $\vec{S}_1$, and $\vec{S}_{1'}$ are aligned parallel to each other, but antiparallel to $\vec{S}_2$ and $\vec{S}_{2'}$. 
However the strong deviation (especially for $\vec{S}_3$, $\vec{S}_2$ and $\vec{S}_{2'}$) from their maximum (classical) value of $\frac12$ indicates that this state has strong QM corrections, as can be directly seen from the explicit form of the wavefunction (not shown). This feature will be explained on the basis of the strong coupling description developed below.     

\begin{figure}
\includegraphics[width=0.95\linewidth]{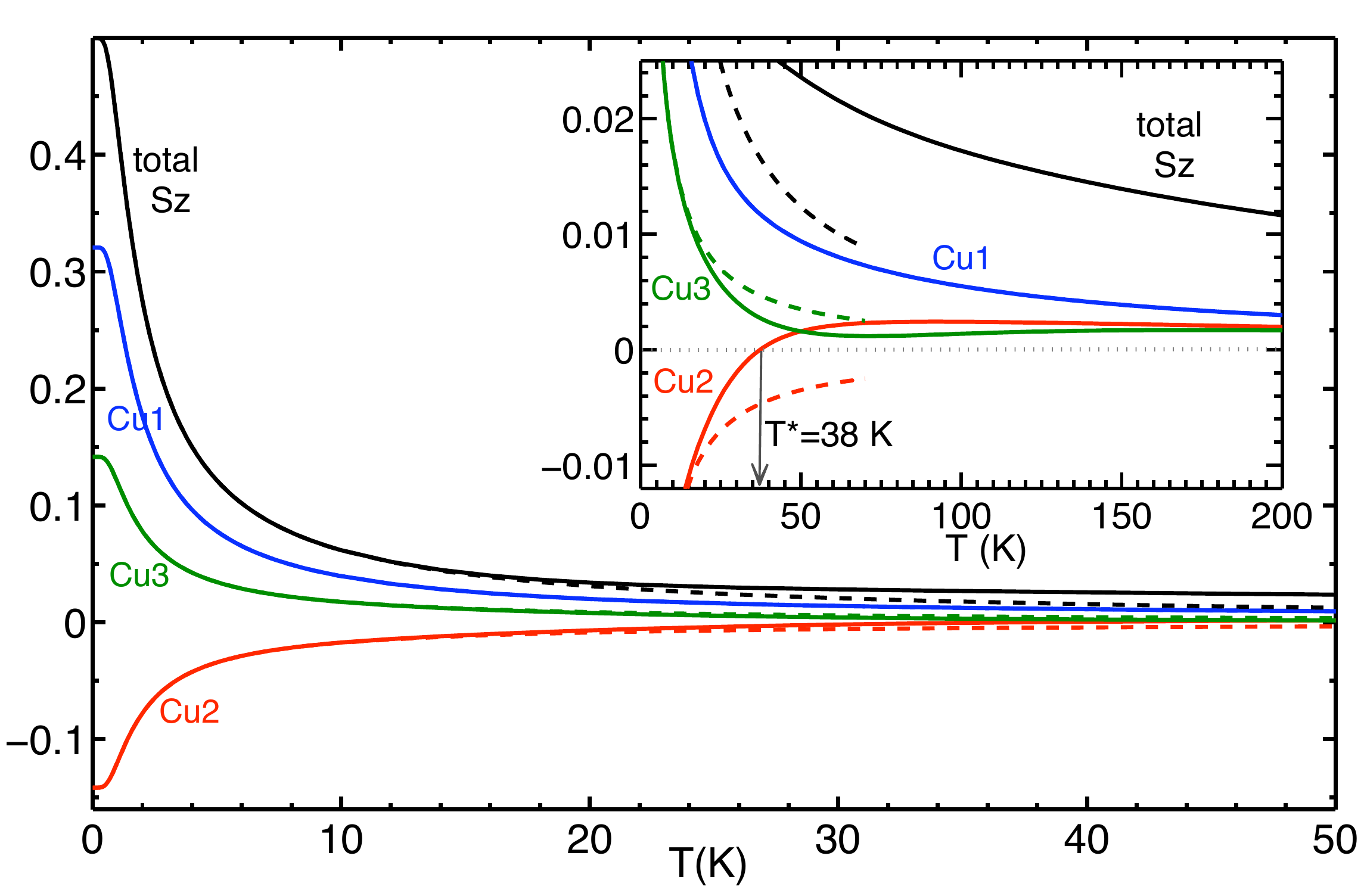}
\caption{\label{fig:localMomentsvsT}
(Color online) Temperature dependence of the local moments of the three inequivalent Cu sites in \cunipa\, as obtained by exact diagonalization using our {\it ab initio} exchange coupling parameters and a field of $\mu_0H=1.6608$~T. The inset shows the same results in a different scale, so that the deviation from the low-$T$ behavior of Eq.~(\ref{eq:ExactMoments}), shown by the dashed lines, is better visible. Also, $T^\ast\simeq 38$ K denotes the characteristic temperature where the Cu2 moments change sign. }
\end{figure}

The deviation of the local moments from Eq.~(\ref{eq:ExactMoments}) at high temperatures is shown in Fig.~\ref{fig:localMomentsvsT} at a field of $1.6608$ T. We note in particular the non-monotonic temperature dependence of the magnitude of the Cu2 moments, which was discussed in Sec.~\ref{sec:nmrLineWidth} above in relation to the NMR lineshape.

To further probe the nature of the GS, we consider the strength of various spin-spin correlations 
$e_{ij}\!\equiv\!\langle\vec{S}_i\cdot\vec{S}_j\rangle_0$. We find 
$e_{31}\!=\!e_{31'}\simeq +0.2122$, 
$e_{32}\!=\!e_{32'}\simeq -0.4810$, 
$e_{12}\!=\!e_{1'2'}\simeq -0.4299$, 
$e_{12'}\!=\!e_{1'2}\simeq -0.2858$, 
and $e_{11'}\!=\!e_{22'}\simeq +0.2345$. 
These values corroborate the above QM ferrimagnetic picture for the GS, and in addition highlight that the bonds 
with the largest exchange coupling $J_{23}$ (i.e., the bonds $3$-$2$ and $3$-$2'$) exhibit also the strongest correlations. 
Moreover, using the above correlations and the values of the exchange couplings, one finds that the $J_{23}$-bonds 
contribute about 72\% of the total GS energy $E_0\simeq -1.3289 J_{23}$.  

We should remark here that the above result $e_{11'}=e_{22'}$ does not arise from the symmetry of the Hamiltonian, but it is an exact property of the particular GS only (see  explanation below).  

{\it Effective g-tensors.}
To make contact with our ESR findings in Sec.~\ref{sec:esr} we calculate the effective g-tensors of the GS and the first excited state in terms of the individual g-tensors of the three inequivalent Cu sites. 
To this end, one needs the equivalent spin operators within each multiplet in order to rewrite the Zeeman Hamiltonian as 
\be
\sum_{i} \vec{B}\cdot\vec{g}_i\cdot\vec{S}_i \mapsto \vec{B}\cdot\vec{g}_{\text{eff}}\cdot\vec{S} ~.
\ee
For the GS, the equivalent operators can be extracted from Eq.~\ref{eqn:GSlocalmoments} above, which yields  
\be
\vec{g}_{\text{eff, GS}} =  +0.0708~\vec{g}_3 -0.1415~\vec{g}_2 + 0.3207~\vec{g}_1~.
\ee
So the deviation from the isotropic $g=2$ value is coming mainly from that of $\vec{g}_1$. 

For the first excited state, our exact wavefunctions give the equivalent operators 
\be\label{eqn:1stExclocalmoments}
\begin{array}{c}
S_3^z \to -\frac{1}{3}~\vec{S},\quad 
S_2^z, S_{2'}^z \to +0.1647~\vec{S},\\
S_1^z, S_{1'}^z \to +0.002~\vec{S}. 
\end{array}
\ee
leading to  
\be
\vec{g}_{\text{eff, 1st exc.}} =  -\frac{1}{3}~\vec{g}_3 + 0.3294~\vec{g}_2 + 0.004~\vec{g}_1~.
\ee
So here the deviation from the isotropic $g=2$ value is mainly governed by that of $\vec{g}_2$ and $\vec{g}_3$. 
The contribution from $\vec{g}_1$ is extremely small because in the first excited doublet the pair of $\vec{S}_1$ and $\vec{S}_{1'}$ forms almost a singlet (see Table~\ref{tab:ExactSpectrum} and Sec.~\ref{sec:SCE} below). 

These expressions demonstrate why the effective g-tensors of the two ESR lines appearing below 70 K are different. 

\subsection{Strong coupling description}\label{sec:SCE}
We are now going to show that almost all properties of the model at $y\equiv J_{12}=-J_{13}=\frac27$ can be adiabatically traced back -- even on a quantitative level -- to the strong coupling limit $y=0$.
 
{\it Global spectral structure.}
At $y=0$, the spins $\vec{S}_1$ and $\vec{S}_{1'}$ are isolated and are, thus, free to point up or down. 
On the other hand, the spins $\vec{S}_3$, $\vec{S}_2$, and $\vec{S}_{2'}$ form an AFM spin trimer with the Hamiltonian 
\be
\mc{H}_0/J_{23} = \vec{S}_3\cdot (\vec{S}_2+\vec{S}_{2'})=\frac{1}{2}\left( \vec{S}_{322'}^2-\vec{S}_{22'}^2-\vec{S}_3^2 \right) . 
\ee 
Hence the spectrum of $\mc{H}_0$ can be derived analytically in terms of the good quantum numbers 
$S_{22'}$ and $S_{322'}$:
\be
\begin{array}{ccccc}
S_{22'}, & S_{322'}, & \text{parity}, & \text{deg}, & E^{(0)}\\
\hline
1 & \frac{1}{2} & \text{even} & 8 & -J_{23} \\
0 & \frac{1}{2} &  \text{odd} & 8 & 0 \\
1 & \frac{3}{2} &  \text{even} & 16 & +J_{23}/2
\end{array}
\ee
where we have also indicated the parity and the degeneracy ($\text{deg}$) of each unperturbed level. 
The latter takes into account the four possible states of the space of $\vec{S}_1$ and $\vec{S}_{1'}$, and the Zeeman degeneracy. 
So we find three well separated unperturbed manifolds. 
The remaining couplings $\mc{V}\!\equiv\! \mc{H}-\mc{H}_0$ split these manifolds but, as we show below, most of these splittings 
are fairly weak, thus explaining the overall spectral structure at $y=\frac27$ (Table~\ref{tab:ExactSpectrum}). 
 
Since each unperturbed manifold has a well-defined $S_{322'}$, we can find the first-order splitting with the use of the ``equivalent operators'':
\be\label{eq:eqops1}
\vec{S}_3 \to \lambda_3\vec{S}_{322'},~~~~\vec{S}_2 \to \lambda_{2}\vec{S}_{322'},~~~
\vec{S}_{2'} \to \lambda_{2'}\vec{S}_{322'},
\ee 
where the values of $\lambda_{1,2,2'}$ can be derived from the coupling (Clebsh-Gordan) coefficients of the respective angular momenta of each manifold (see below). Using $\lambda_2\!=\!\lambda_{2'}$, which holds in all manifolds, and replacing \eqref{eq:eqops1} in $\mc{V}$, we obtain
\be
\mc{V} \to J_{\text{eff}}~\vec{S}_{322'}\cdot\vec{S}_{11'} = \frac{1}{2}J_{\text{eff}} \left(
\vec{S}^2-\vec{S}_{322'}^2-\vec{S}_{11'}^2\right),
\ee
where $J_{\text{eff}}\!=\!(\lambda_2-\lambda_3) y$. Thus, to lowest order, the perturbation gives rise to an effective exchange coupling between $\vec{S}_{11'}$ and $\vec{S}_{322'}$. 

For the lowest eight-dimensional manifold, we couple the spin $S_3\!=\!\frac12$ to the spin $S_{22'}\!=\!1$ to get an $S_{322'}\!=\!\frac12$ object, which gives (see Appendix \ref{app:GGcoefs}): $\lambda_3\!=\!-\frac13$, $\lambda_2\!=\!\lambda_{2'}\!=\!+\frac23$, and thus $J_{\text{eff}}\!=\!y\!>\!0$.  The resulting correction to the lowest unperturbed manifold reads: 
\be
\begin{array}{ccccccc}
S_{22'}, & S_{322'}, & S_{11'}, & S, & \text{parity}~p, & \text{deg}, & \Delta E^{(1)}\\
\hline
1 & \frac{1}{2} & 1 & \frac{1}{2} &  \text{even} & 2 & -y \\
1 & \frac{1}{2} & 0 & \frac{1}{2} &  \text{odd} & 2 &  0 \\
1 & \frac{1}{2} & 1 & \frac{3}{2} &  \text{even} & 4 & +y/2
\end{array}
\ee

\begin{figure*}
\includegraphics[width=0.85\linewidth]{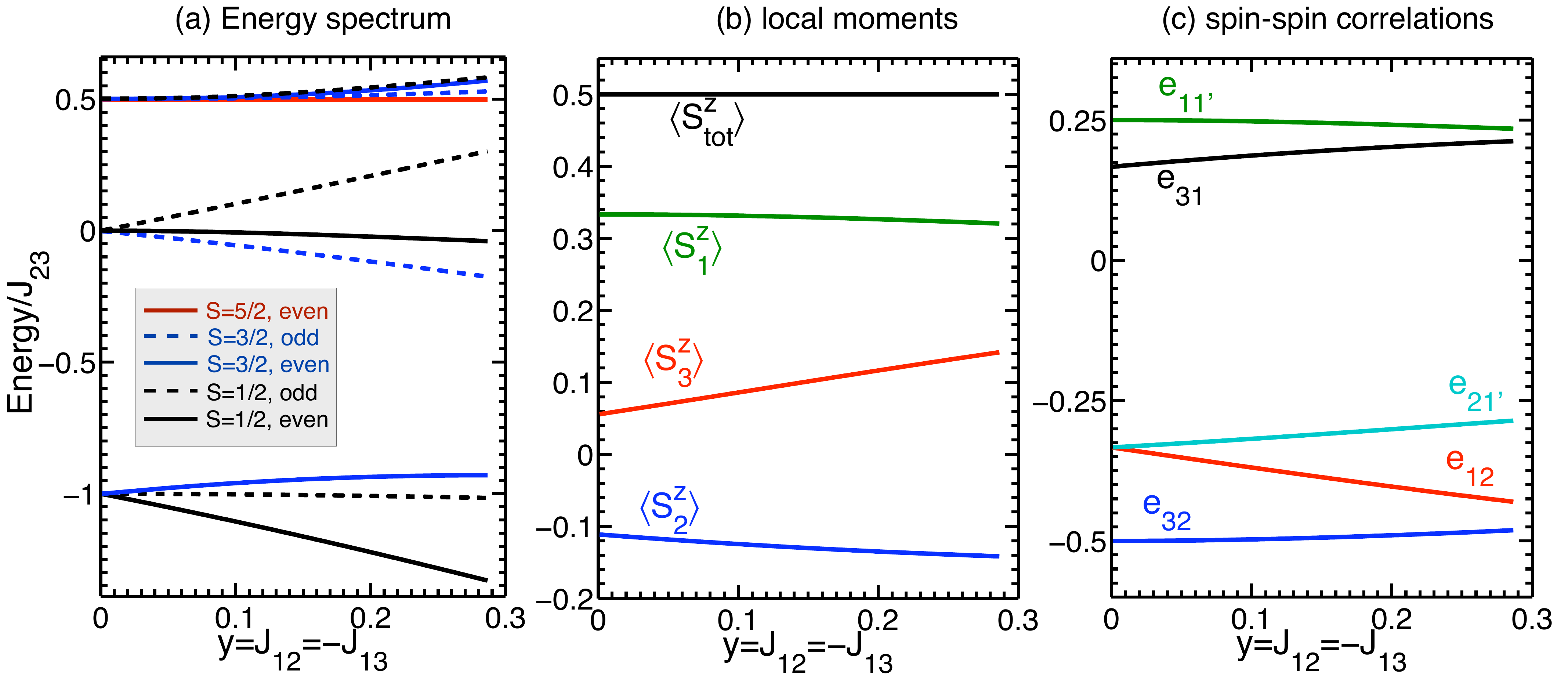}
\caption{(Color online) Accuracy of the strong-coupling description. 
Evolution of the energy spectrum (a), the local moments (b), 
and the spin-spin correlations $e_{ij}\!=\!\langle\vec{S}_i\cdot\vec{S}_j\rangle$ (c) from the strong coupling limit $y\!=\!0$ 
up to $y\!=\!\frac27$. The energies are given in units of $J_{23}$.
}\label{fig:SpectrumMomentsCorrsvsy}
\end{figure*}

Similarly, the second unperturbed manifold has $S_{22'}\!=\!0$ and thus $\lambda_2\!=\!\lambda_{2'}\!=\!0$ and 
$\lambda_3\!=\!1$, giving $J_{\text{eff}}\!=\!-y$, which is now FM. 
So the first-order splitting of this manifold has the opposite structure from that of the lowest manifold, namely:  
\be
\begin{array}{ccccccc}
S_{22'}, & S_{322'}, & S_{11'}, & S, & \text{parity}~p, & \text{deg}, & \Delta E^{(1)}\\
\hline
0 & \frac{1}{2} & 1 & \frac{3}{2} &  \text{odd} & 4 &  -y/2 \\
0 & \frac{1}{2} & 0 & \frac{1}{2} & \text{even} & 2 & 0 \\
0 & \frac{1}{2} & 1 & \frac{1}{2} &  \text{odd} & 2 & +y
\end{array}
\ee
The third manifold has $S_{322'}\!=\!\frac32$ and thus $\lambda_3\!=\!\lambda_2\!=\!\lambda_{2'}\!=\!\frac13$, 
which in turn gives $J_{\text{eff}}\!=\!0$. Thus, the third unperturbed manifold remains intact in lowest order, 
i.e., all 16 states have $\Delta E^{(1)}\!=\!0$:
\be
\begin{array}{ccccccc}
S_{22'}, & S_{322'}, & S_{11'}, & S, & \text{parity}~p, & \text{deg}, & \Delta E^{(1)}\\
\hline
1 & \frac{3}{2} & 1 & \frac{1}{2} &  \text{even} & 2 &  0 \\
1 & \frac{3}{2} & 1 & \frac{3}{2} &  \text{even} & 2 & 0 \\
1 & \frac{3}{2} & 1 & \frac{5}{2} &  \text{even} & 4 & 0 \\
1 & \frac{3}{2} & 0 & \frac{3}{2} &  \text{odd}  & 4 & 0
\end{array}
\ee
In particular, the energy of the state with the maximum spin ($S=\frac52$) remains equal to $J_{23}/2$ in all orders of perturbation theory, as can be easily checked, e.g., for its maximum polarized $M=\frac52$ portion, which is a trivial eigenstate of $\mc{H}$.

Up to this lowest order, the level-crossing fields $H_{c1}$ and $H_{c2}$ are given by
\be
g\mu_B H_{c1}/k_B \simeq 3y/2 , ~~~ 
g\mu_B H_{c2}/k_B \simeq  (3J_{23}-y)/2 . 
\ee
With $J_{23}=217$ K, $y=2J_{23}/7$, and the average $g\simeq 2.2$, these numbers give: 
$H_{c1}\simeq 54$ T, and $H_{c2}\simeq 170$ T, which are close to the exact numbers at $y=\frac27$ given above.   

Figure~\ref{fig:SpectrumMomentsCorrsvsy} shows the evolution of the spectrum from $y\!=\!0$ up to $y\!=\!\frac27$,
and demonstrates that the above first-order spectral structure reproduces quantitatively the spectrum at $y\!=\!\frac27$. 

We should add here that $S_{22'}$, $S_{322'}$, and $S_{11'}$ do not remain good quantum numbers 
in higher orders of the perturbative expansion, which is expected, since the corresponding angular momenta 
are not conserved quantities under the full Hamiltonian. Still, their expectation values reported in Table~\ref{tab:ExactSpectrum} 
are very close to the above strong-coupling values. 

As a final remark, we note that one may use the above basis provided by the lowest-order theory and the 
good quantum numbers $S$, $M$, and $p$ to find the invariant subspaces of $\mc{H}$. 
Most of the subspaces turn out to be two-dimensional, 
except for the space with $S=M=\frac12$ and $p=\,$even (in which the GS belongs), which is three-dimensional. 
Here, it may be readily checked that each of the three states has $S_{11'}\!=\!S_{22'}$, thus explaining   
the equality $e_{11'}\!=\!e_{22'}$ discussed above. The excited states do not share this property, so at higher temperatures 
the two correlations deviate from each other. 

{\it GS properties.}
Let us now return to the nature of the GS. According to the above lowest order result, the GS is a total $S=\frac12$ doublet 
with even parity and total energy $E^{(1)}\!=\!-J_{23}-J_{12}$. To understand the nature of this state, we note that it arises from the coupling of a spin $s\!=\!\frac12$ (here $\vec{S}_{322'}$) and a spin $s=1$ (here $\vec{S}_{11'}$) to a total $S\!=\!\frac12$ state. So we may again use the ``equivalent operators'' 
\be
\vec{S}_{322'} \to -\frac{1}{3} \vec{S},~~~\vec{S}_1, \vec{S}_{1'}  \to +\frac{2}{3} \vec{S} ,
\ee 
which in conjunction with the ones relating $\vec{S}_1$, $\vec{S}_3$, and $\vec{S}_5$ to $\vec{S}_{322'}$ (see Eq.~(\ref{eq:eqops1}) above) gives the central relations:  
\be\label{eq:eqops2}
\begin{array}{c}
\vec{S}_3 \to +\frac{1}{9}\vec{S}, \\
\vec{S}_2, \vec{S}_{2'} \to -\frac{2}{9} \vec{S},~~~
\vec{S}_1, \vec{S}_{1'} \to +\frac{2}{3} \vec{S}.
\end{array}
\ee 
This equation provides a very clear explanation for the peculiar local moment distribution discussed above, 
since, when the total $S=\frac12$ moment is saturated (by applying a small field), this equation gives: 
\be
\begin{array}{c}
\langle S_3^z\rangle = +\frac{1}{18}, \\
\langle S_2^z\rangle=\langle S_{2'}^z\rangle =-\frac{1}{9},~~
\langle S_1^z\rangle=\langle S_{1'}^z\rangle = +\frac{1}{3}.
\end{array}
\ee
The same numbers can be extracted directly from the explicit form of the GS wavefunction, which for the $M=+\frac12$ portion reads: 
\bea\label{eq:GSsclimit}
|\!\Uparrow\rangle \!&=&\! 
-\frac{2}{3} |\!\uparrow\rangle_3 |t_1\rangle_{11'}   |t_{-1}\rangle_{22'}
+\frac{1}{3}|\!\uparrow\rangle_3 |t_0\rangle_{11'} |t_0\rangle_{22'} \nonumber\\
&+&\frac{\sqrt{2}}{3} |\!\downarrow\rangle_3 
\big(  |t_1\rangle_{11'} |t_0\rangle_{22'} 
- |t_0\rangle_{11'} |t_1\rangle_{22'}  \big),
\eea
where 
$|t_1\rangle_{ij}\!=\!|\!\!\!\uparrow_i\!\uparrow_j\rangle$, 
$|t_{-1}\rangle_{ij}\!=\!|\!\!\!\downarrow_i\!\downarrow_j\rangle$, 
and 
$|t_0\rangle_{ij}\!=\!\left( |\!\uparrow_i\!\downarrow_j\rangle + |\!\downarrow_i\!\uparrow_j\rangle \right)/\sqrt{2}$.  
The leading term of Eq.~(\ref{eq:GSsclimit})) is the classical ferrimagnetic configuration. However, the overall form of the wavefunction demonstrates explicitly the presence of strong QM corrections to the GS.  

The strong coupling limit reproduces the ferrimagnetic arrangement of the local moments and even their strengths, except for $\vec{S}_3$ which has almost tripled its moment at $y=\frac27$. This is further demonstrated in Fig.~\ref{fig:SpectrumMomentsCorrsvsy}(b), which shows the evolution of the local moments from $y\!=\!0$ up to $y\!=\!\frac27$. 
All moments, except $\langle S_3^z\rangle$, exhibit a small slope, i.e., higher-order corrections are weak. 

In a similar fashion, the strong coupling prediction for the spin-spin correlations are: 
$e_{31}\!=\!e_{31'}\!=\!\frac16$, 
$e_{32}\!=\!e_{32'}\!=\!-\frac12$, 
$e_{12}\!=\!e_{1'2'}\!=\!-\frac13$, 
$e_{12'}\!=\!e_{1'2}\!=\!-\frac13$, 
and $e_{11'}\!=\!e_{22'}\!=\!\frac14$, 
which are in close agreement with the exact values for $y\!=\!\frac27$ given earlier. This is again demonstrated in Fig.~\ref{fig:SpectrumMomentsCorrsvsy}(c) that shows the evolution of the spin-spin correlation strengths from $y\!=\!0$ up to $y\!=\!\frac27$. All correlations exhibit weak corrections, except $e_{12}$ which shows a somewhat larger slope in $y$. 
 
\section{Discussion and Summary}
\label{sec:discussion}
The \cunipa\ molecule comprises two non-frustrated spin triangles with two AFM couplings and one FM coupling each (Fig.~\ref{fig:structure}, left). This coupling regime is rather unusual and deserves a further analysis with respect to the underlying crystal structure and interacting orbitals. Three leading couplings -- $J_{12},J_{13}$, and $J_{23}$ -- represent the Cu--O--Cu superexchange running via the $\mu_3$-O atom in the middle of the Cu$_3$ triangle. According to Goodenough-Kanamori rules, high values of the bridging angle should lead to an AFM interaction, whereas low bridging angles close to 90$^{\circ}$ favor FM couplings. Our results for \cunipa\ (Table~\ref{tab:couplings}) follow this general trend. However, a closer examination pinpoints two peculiarities of this compound.

First, the coupling remains FM for the bridging angle of $107.9^{\circ}$, although the standard and commonly accepted threshold value of the FM-AFM crossover is slightly below $100^{\circ}$.\cite{braden1996} Second, $J_{12}^{\FM}$ is much larger than $J_{13}^{\FM}$, even though the respective bridging angle is also larger and should lead to a smaller FM contribution. Both peculiarities should be traced back to the twisted configuration of the interacting CuO$_4$ plaquettes (Fig.~\ref{fig:structure}, left). The systematic work on the angular dependence of the exchange coupling is usually restricted to systems with two plaquettes lying in the same plane or only weakly twisted (the dihedral angle between the planes is close to 180~deg). The Cu$_5$ molecule represents an opposite limit of strongly twisted CuO$_4$ plaquettes, with the dihedral angles of 91.8~deg ($J_{12}$), 125.2~deg ($J_{13}$), and 99.9~deg ($J_{23}$) between the CuO$_4$ planes.\footnote{Here, we averaged four Cu--O vectors of each plaquette, because Cu and O atoms do not form a single plane owing to the low crystallographic symmetry of the compound.}

The large FM contribution to $J_{12}$ ($J_{12}^{\FM}=-134$~K) can be ascribed to the nearly orthogonal configuration of the interacting CuO$_4$ plaquettes. However, the mechanism of this FM interaction is yet to be determined. The large FM coupling for the bridging angle of 90~deg is generally understood as the Hund's coupling on the oxygen site\cite{mazurenko2007} or the direct FM exchange between Cu and O.\cite{kuzian2012} Both mechanisms depend solely on the Cu--O--Cu angle and should be rather insensitive to the mutual orientation of the CuO$_4$ plaquettes. Therefore, other effects, such as the direct Cu--Cu exchange and the Cu--O--Cu interaction involving fully filled Cu $3d$ orbitals, may be operative. Experimental information on the magnetic exchange between the strongly twisted CuO$_4$ plaquettes remains scarce and somewhat unsystematic. The Cu--O--Cu angle of 104.5~deg combined with the sizable twisting give rise to a FM nearest-neighbor interaction in the kagome material kapellasite Cu$_3$Zn(OH)$_6$Cl$_2$,\cite{fak2012} whereas a similar geometry with the Cu--O--Cu angle of 107.6~deg results in an overall AFM coupling in dioptase, Cu$_6$Si$_6$O$_{18}\cdot 6$H$_2$O.\cite{janson2010} 

We also note that \cunipa\ brings a fresh perspective on magnetostructural correlations in Cu$_3$ triangular molecules, where the Cu--X--Cu bridging angle was previously considered the key geometrical parameter. As long as the ligand atom X lies in the Cu$_3$ plane, the bridging angles remain close to 120~deg and should generally lead to the AFM exchange. The FM exchange would only be possible when the ligand atoms are shifted out of the plane, as in the Cl- and Br-containing Cu$_3$ molecules where the Cu--X--Cu angles are below 90~deg.\cite{boca2003} The twisting of the CuX$_4$ plaquettes provides another opportunity for creating the FM exchange and, moreover, for introducing it selectively. Surprisingly, organic ligands have only a weak effect on the magnetic interactions. Although carboxyl groups (COO$^-$) provide an additional superexchange pathway between Cu1 and Cu3 (Fig.~\ref{fig:structure}, left), their molecular orbitals do not influence the Cu $d_{x^2-y^2}$ Wannier functions (Fig.~\ref{fig:wannier}). For example, the FM contribution to $J_{12}$ exceeds that to $J_{13}$. This demonstrates that the FM exchange is basically unrelated to the organic ligands. It should be understood as a joint effect of the the low Cu--O--Cu angle and twisting.

The combination of FM and AFM couplings also has an important effect on the magnetic GS of \cunipa. A regular spin triangle entails the four-fold degenerate GS that further splits into two close-lying doublets by virtue of anisotropy,\cite{chaboussant2002,*furukawa2007} residual interactions between the triangles,\cite{ioannis2005} or a marginal distortion of the triangle.\cite{choi2006,*choi2008} In \cunipa, there is only one doublet state, which is separated by about 68~K from the first excited state (see Table~\ref{tabExactSectrum}). According to Kramer's theorem, in zero field the degeneracy of this state can not be lifted, hence the tunnel splitting is exactly zero, and no magnetization steps (Landau-Zener-St\"uckelberg transitions) should occur, in contrast to, e.g., the V$_6$ molecule.\cite{ioannis2005} Therefore, broad butterfly hysteresis effects but without tunneling are expected in the magnetization process of \cunipa.

Of particular interest is our experimental finding of the enhanced $^1$H nuclear spin-lattice relaxation rate $1/T_1$ at 
a characteristic temperature slightly below the spin gap ($T\simeq 40$ K). While such an enhancement has been repeatedly found in numerous AFM homometallic\cite{baek2004} and heterometallic\cite{amiri2010} rings of spins $S>\frac12$, it is very rare for $S\!=\!\frac{1}{2}$ and has been observed only in molecules with a high-spin GS.\cite{lascialfari1998,carretta2006}  
The origin of the peak is likely the same in both cases, namely, the slowing down of the phonon-driven magnetization fluctuations.\cite{baek2004,santini2005,ioannisPRB2007,ioannisPRB2009} However, the sparse excitation spectrum and the presence of nonequivalent Cu sites with different local magnetization render \cunipa\ dissimilar to typical homometallic rings. We argue that in \cunipa\ inter-multiplet Orbach processes make the dominant contribution to the spin-lattice relaxation process, as in the heterometallic ring Cr$_7$Ni at very low temperatures.\cite{bianchi2010}    

The use of molecular magnets in quantum computing is severely restricted by their short coherence time. Coherence times on the order of 100~$\mu$s could be achieved in, e.g., V$_{15}$ by arranging magnetic molecules in a self-assembled layer formed by an organic surfactant.\cite{bertaina08} This method is fundamentally similar to the formation of the metal-organic framework in \cunipa. Therefore, it may be interesting to study the decoherence process in this molecular magnet and, more generally, explore the role of organic bridges in the decoherence process.

Finally, we would like to note that only a few examples of magnetic pentamers based on spin-$\frac12$ ions have been reported so far. The [Cu($\mu$-L)$_3$]$_2$Cu$_3$($\mu_3$-OH)(PF$_6)_3\cdot 5$H$_2$O ($L^-=3,5$-bis(2-pyridil)pyrazolate) shows a somewhat similar phenomenology with the spin $S\!=\!\frac12$ GS and the respective magnetization plateau ranging up to 30~T.\cite{ishikawa2010} However, the geometrical structure of this compound is a trigonal bipyramid that is notably different from the hourglass shape of the magnetic cluster in \cunipa. The magnetic pentamer based on Cu$_5$(OH)$_4$(H$_2$O)$_2$(A-$\alpha$-SiW$_9$O$_{33})_2]$ is more similar to our case and also reveals the spin $S\!=\!\frac12$ GS.\cite{bi2004,nellutla2005} Nevertheless, the equivalence of $J_{23}$ and $J_{13}$, as well as sizable long-range couplings, distinguish its spectrum and magnetic properties from that of \cunipa.

In summary, we have studied thermodynamic properties, spin dynamics, microscopic magnetic model, and energy spectrum of the spin-$\frac12$ Cu$_5$ pentamer in \cunipa. This magnetic molecule has an hourglass shape with two AFM couplings and one FM coupling on each of the two triangles that form the non-frustrated pentamer. In zero field, the ground state of \cunipa\ is a doublet with the total spin of $S=\frac12$. This ground state is separated from the first excited state by an energy gap of $\Delta\simeq 68$~K. Our results evidence a highly inhomogeneous distribution of magnetization over rhe nonequivalent Cu sites according to the quantum nature of the magnetic ground state. The maximum in the spin-lattice relaxation rate $1/T_1$ is very rare among molecular magnets with spin-$\frac12$ ions and a low-spin ground state.

\begin{acknowledgments}
RN was funded by MPG-DST (Max Planck Gesellschaft, Germany and Department of Science and Technology, India) fellowship. 
AT was supported by the Alexander von Humboldt Foundation and the Mobilitas program of the ESF (grant MTT77). OJ acknowledges partial support of the Mobilitas grant MJD447. IR was funded by the Deutsche Forschungsgemeinschaft (DFG) under the Emmy-Noether program. The high-field magnetization measurements were supported by EuroMagNET~II under the EC contract 228043. 
Finally, we would like to thank M. Belesi for fruitful discussions.
\end{acknowledgments}

\appendix
\section{Coupling of a spin 1 to a spin 1/2 object and ``equivalent operators''}\label{app:GGcoefs}
Here we work out the coupling of two spins, one with spin $S_a=1/2$ and the other one with $S_{bc}=1$. For the latter, we shall imagine that there are two spins $S_b=S_c=1/2$ forming a triplet, namely $|t_1\rangle=|\!\uparrow\uparrow\rangle$, $|t_{-1}\rangle=|\!\downarrow\downarrow\rangle$, and $|t_0\rangle=\frac{1}{\sqrt{2}}(|\!\uparrow\downarrow\rangle+|\!\downarrow\uparrow\rangle)$. 
The states of the system can be labeled by $|S, M\rangle$, where 
$S$ is the total spin (here $S=\frac12$ or $\frac32$), and $M$ is the projection along some axis $\vec{z}$. 
The states with $S=\frac32$ are given by:
\bea
&& |\tfrac{3}{2},\tfrac{3}{2}\rangle=|\!\uparrow\rangle_a\otimes|t_1\rangle_{bc} , \nonumber\\
&& |\tfrac{3}{2},\tfrac{1}{2}\rangle=\frac{1}{\sqrt{3}} \left(
|\!\downarrow\rangle_a\otimes|t_1\rangle_{bc}
+ \sqrt{2} |\!\uparrow\rangle_a\otimes|t_0\rangle_{bc} \right) ,  \nonumber\\
&& |\tfrac{3}{2},-\tfrac{1}{2}\rangle=\frac{1}{\sqrt{3}}\left( |\!\uparrow\rangle_a\otimes|t_{-1}\rangle_{bc}
+\sqrt{2} |\!\downarrow\rangle_a\otimes|t_0\rangle_{bc} \right) , \nonumber\\
&& |\tfrac{3}{2},-\tfrac{3}{2}\rangle=|\!\downarrow\rangle_a\otimes|t_{-1}\rangle_{bc} .\nonumber
\eea
On the other hand, the two states of the $S=\frac12$ doublet read: 
\bea
&& |\tfrac{1}{2},\tfrac{1}{2}\rangle= \frac{1}{\sqrt{3}}\left(
 |\!\uparrow\rangle_a\otimes|t_0\rangle_{bc}-\sqrt{2} |\!\downarrow\rangle_a\otimes|t_1\rangle_{bc}  \right) , \nonumber\\
&& |\tfrac{1}{2},-\tfrac{1}{2}\rangle=\frac{1}{\sqrt{3}} \left(
-|\!\downarrow\rangle_a\otimes|t_0\rangle_{bc}+\sqrt{2} |\!\uparrow\rangle_a\otimes|t_{-1}\rangle_{bc}  \right) .\nonumber
\eea
Using the above explicit relations, it is straightforward to find out the equivalent operators 
within the $2\times 2$ manifold of the $S=\frac12$ doublet. 
We find: 
\be
\vec{S}_a \to -\frac{1}{3}\vec{S}, ~~~\vec{S}_b \to +\frac{2}{3}\vec{S}, ~~~\vec{S}_c \to +\frac{2}{3}\vec{S} .
\ee

\section{Structural data}\label{app:B}
Table \ref{tab:structuraldata} shows the equilibrium positions of the 17 hydrogen nuclei in the structure. 

\begin{table}[!ht]
\caption{Relaxed hydrogen positions in Cu$_5$-NIPA. Last column lists atoms bonded to the given hydrogen atom. 
The notation of atoms follows Ref.~\onlinecite{liu2011}, $a=10.8303$~\AA, $b=11.4692$~\AA, and $c=11.5697$~\AA.
}\label{tab:structuraldata}
\begin{ruledtabular}
\begin{tabular}{ccccc}
      & $x/a$  & $y/b$  & $z/c$  &       \\
 H2   & 0.0341 & 0.9481 & 0.3828 & (C2)  \\
 H4   & 0.4295 & 0.6383 & 0.4380 & (C4)  \\
 H6   & 0.4442 & 0.0066 & 0.2381 & (C6)  \\
 H10  & 0.4563 & 0.4564 & 0.2062 & (C10) \\
 H12  & 0.1278 & 0.8023 & 0.0923 & (C12) \\
 H13  & 0.0814 & 0.3482 & 0.1695 & (O13) \\
 H14  & 0.5705 & 0.7834 & 0.9955 & (C14) \\
 H14A & 0.1105 & 0.1813 & 0.9704 & (O14) \\
 H14B & 0.0012 & 0.2993 & 0.9133 & (O14) \\
 H15A & 0.6634 & 0.2163 & 0.2633 & (O15) \\
 H15B & 0.8008 & 0.0961 & 0.2947 & (O15) \\
 H16A & 0.9178 & 0.6522 & 0.4705 & (O16) \\
 H16B & 0.8144 & 0.5794 & 0.5639 & (O16) \\
 H17A & 0.3996 & 0.2783 & 0.2640 & (O17) \\
 H17B & 0.4504 & 0.1667 & 0.3606 & (O17) \\
 H18A & 0.0954 & 0.1403 & 0.6165 & (O18) \\
 H18B & 0.1939 & 0.1190 & 0.6921 & (O18) \\
\end{tabular}
\end{ruledtabular}
\end{table}

%

\end{document}